%% file: main.tex
\documentclass[acmtog,authorversion]{acmart}
\usepackage{subfigure}
\usepackage{siunitx}
\usepackage{cleveref}
\usepackage[utf8]{inputenc}
\usepackage{framed}
\usepackage{listings}
\usepackage{tabularx}
\usepackage{multirow} 
\usepackage{xcolor} 
\usepackage{float} 
\usepackage{caption}

\input{sec/_template_args.tex}

\begin{document}
\title{{Social Agent: Mastering Dyadic Nonverbal Behavior Generation via Conversational LLM Agents}}

\input{sec/0_authors.tex}
\input{sec/0_abstract}

%
%

 
%
%

\begin{teaserfigure}
  \centering
  \includegraphics[width=\textwidth]{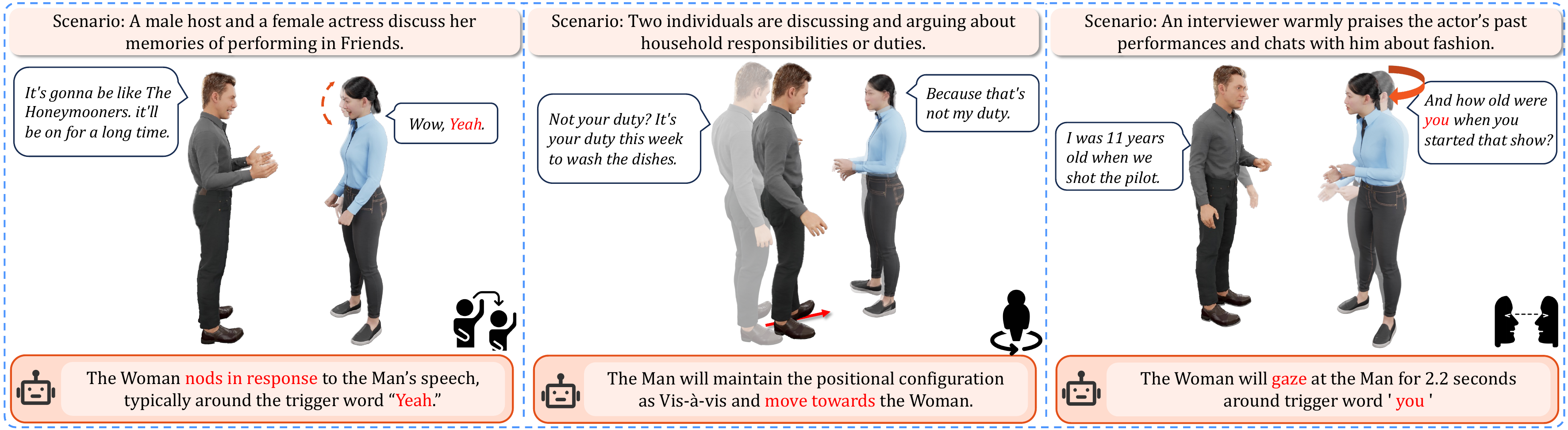}
  \caption{Our system generates natural and context-aware dyadic nonverbal behaviors via LLM-guided interaction control and dual-person gesture synthesis.}
  \Description{}
  \label{fig:teaser}
\end{teaserfigure}

\maketitle

\input{sec/1_intro}
\input{sec/2_related_work}

\input{sec/3_approach}

\input{sec/4_experiment}

\input{sec/5_conclusion}

\begin{acks}
  We thank the anonymous reviewers for their constructive comments. This work was supported in part by National Key R\&D Program of China 2022ZD0160803.
\end{acks}

\bibliographystyle{ACM-Reference-Format} 
\bibliography{main}
\newpage

\newpage
\input{sec/appendix}


\end{document}

%% file: sec/_template_args.tex
\AtBeginDocument{%
  \providecommand\BibTeX{{%
    \normalfont B\kern-0.5em{\scshape i\kern-0.25em b}\kern-0.8em\TeX}}}

\copyrightyear{2025}
\acmYear{2025}
\setcopyright{acmlicensed}
\acmConference[SA Conference Papers '25]{SIGGRAPH Asia 2025 Conference Papers}{December 15--18, 2025}{Hong Kong, China}
\acmBooktitle{SIGGRAPH Asia 2025 Conference Papers (SA Conference Papers '25), December 15--18, 2025, Hong Kong, China}
\acmDOI{10.1145/3757377.3763879}
\acmISBN{979-8-4007-2137-3/2025/12}

\acmSubmissionID{0}



%% file: sec/0_authors.tex
\author{Zeyi Zhang}
\email{illusence1@gmail.com}
\affiliation{%
    \institution{School of Intelligence Science and Technology, Peking University}
    \country{China}
}

\author{Yanju Zhou}
\email{yanjuzhou331@gmail.com}
\affiliation{%
    \institution{Yuanpei College, Peking University}
    \country{China}
}

\author{Heyuan Yao}
\email{heyuanyao@pku.edu.cn}
\affiliation{%
    \institution{School of Computer Science, Peking University}
    \country{China}
}

\author{Tenglong Ao}
\email{aubrey.tenglong.ao@gmail.com}
\affiliation{%
    \institution{School of Computer Science, Peking University}
    \country{China}
}

\author{Xiaohang Zhan}
\email{xiaohangzhan@outlook.com}
\orcid{0000-0003-2136-7592}
\affiliation{%
    \institution{Tencent}
    \country{China}
}

\author{Libin Liu}
\authornote{corresponding author}
\email{libin.liu@pku.edu.cn}
\orcid{0000-0003-2280-6817}
\affiliation{%
  \institution{State Key Laboratory of General Artificial Intelligence, Peking University}
  \country{China}
}


%% file: sec/0_abstract.tex
\begin{abstract}
We present Social Agent, a novel framework for synthesizing realistic and contextually appropriate co-speech nonverbal behaviors in dyadic conversations. In this framework, we develop an agentic system driven by a Large Language Model (LLM) to direct the conversation flow and determine appropriate interactive behaviors for both participants. Additionally, we propose a novel dual-person gesture generation model based on an auto-regressive diffusion model, which synthesizes coordinated motions from speech signals. The output of the agentic system is translated into high-level guidance for the gesture generator, resulting in realistic movement at both the behavioral and motion levels. Furthermore, the agentic system periodically examines the movements of interlocutors and infers their intentions, forming a continuous feedback loop that enables dynamic and responsive interactions between the two participants. User studies and quantitative evaluations show that our model significantly improves the quality of dyadic interactions, producing natural, synchronized nonverbal behaviors. 
We will release the code and prompts for academic research.
\end{abstract}

%% file: sec/1_intro.tex
\section{Introduction}

Nonverbal behaviors are a crucial and indispensable part of human communication. They often convey nuanced social signals, such as emotions, attitudes, and social relationships, at multiple scales \cite{hall1973silent, knapp1972nonverbal}. In dyadic conversations, for instance, interlocutors naturally maintain a certain spatial distance, reflecting their social relationships and level of familiarity. At a broader behavioral scale, eye contact is a well-observed behavior when interlocutors seek engagement and attentiveness. 
Moreover, interlocutors often exhibit gesture synchrony, which encompasses both the \emph{chameleon effect}—spontaneous imitation of the partner’s gestures \cite{chartrand1999chameleon}—and feedback behaviors such as nodding.
On a more granular level, interlocutors often accompany their speech with body gestures, reinforcing or complementing verbal messages to enhance communication. These nonverbal cues typically emerge instinctively and unconsciously, without individuals being fully aware of them, offering an authentic and unfiltered glimpse into human intent and emotion.

Recent advances in deep learning have enabled the data-driven synthesis of single-person behaviors from speech, particularly co-speech gestures and facial expressions \cite{ao2023gesturediffuclip, pan2024s3, zhang2024semanticgesture}. However, it remains challenging to extend these methods to dyadic conversational scenarios to capture the nuanced social dynamics at all scales. The interactions between interlocutors create complex spatial and temporal dependencies in their fine-grained behaviors. 
High-level behaviors such as eye contact, the chameleon effect, and social distancing are sparsely distributed within these finer-grained behaviors. Approaches that rely solely on data and supervised learning \cite{huang2024interact, shi2024ittakestwo,qi2025co3gesturecoherentconcurrentcospeech} tend to overfit to certain dominant fine-grained behaviors in the training data but fail to capture the sparse but critical dyadic social signals.
Meanwhile, nonverbal and social behaviors in human communication have been extensively studied in psychology and linguistics-related fields \cite{hall1973silent, knapp1972nonverbal, chartrand1999chameleon, KENDON196722}.
Our key insight is that the findings from these studies can be leveraged to inform the design of effective control signals for data-driven generators to create dyadic social interactions. However, bridging this abstract, descriptive knowledge with concrete motion data is a non-trivial challenge. It requires a carefully designed synergy between high-level reasoning and a low-level motion synthesis model.

These observations inspire an agentic solution powered by Large Language Models. Unlike a rigid rule-based system, an LLM-powered agent leverages its semantic understanding to dynamically infer social context and apply appropriate behavioral rules, effectively handling the diversity and complexity of human conversation. We argue that an LLM-driven agent, when equipped with the necessary knowledge, can mimic the instinctive process behind human conversational behavior through modular reasoning. It can infer conversational phases and social intentions from the content of the conversation and the current state of the interlocutors, which then guides context-aware motion execution. By embedding this hierarchical reasoning into nonverbal behavior synthesis, we enhance generative models by explicitly modeling the causal links between multiscale social signals and their embodied expressions.

\begin{figure}[t]
  \centering
  \includegraphics[width=\columnwidth]{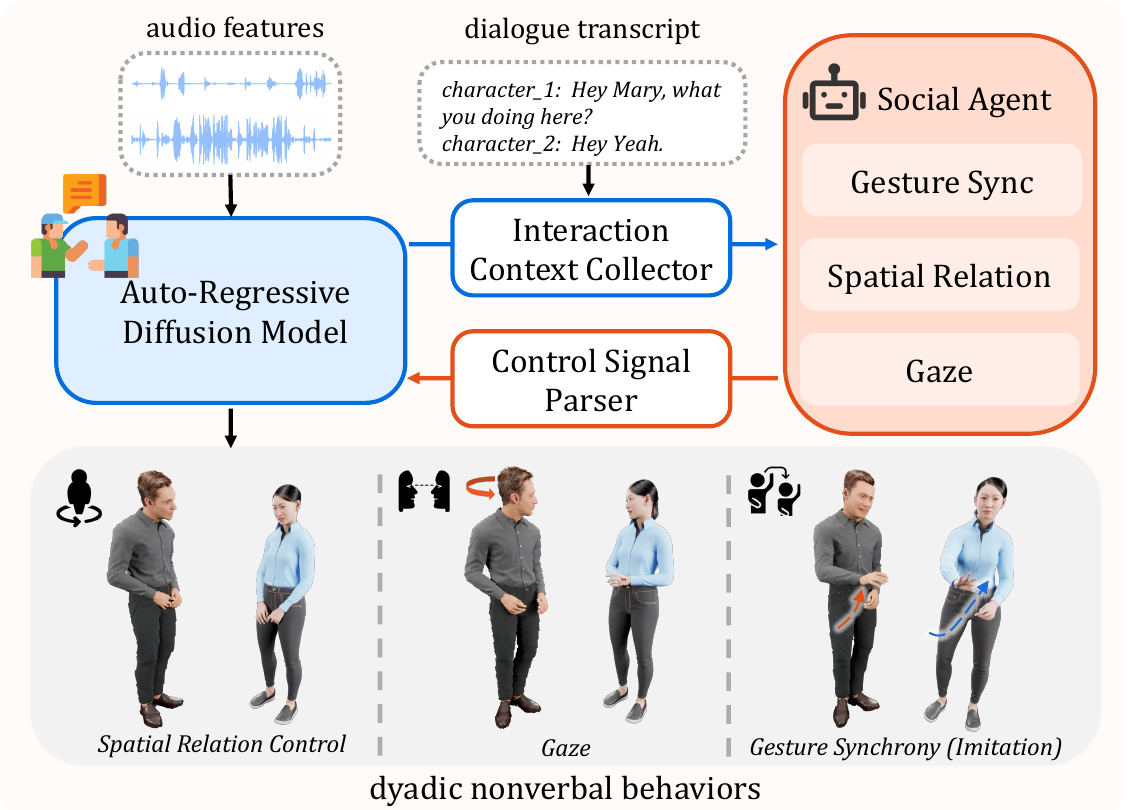}
  \caption{Our framework models dyadic interactions by integrating an autoregressive diffusion model for low-level motion generation with an LLM-based agentic system, Social Agent, for nonverbal behavior analysis. This system continuously analyzes and refines nonverbal behavior cues, dynamically guiding the diffusion model to generate natural interpersonal behaviors such as spatial positioning, gaze contact, and gesture synchrony}.
  \label{fig:overview}
\end{figure}

As shown in \Cref{fig:overview}, we construct our dyadic nonverbal behavior generation system by first designing an auto-regressive diffusion model as a high-quality behavior generator capable of bidirectional and temporally entangled generation. Based on this model, we introduce an agentic framework that acts as a \emph{Director}, named Social Agent System, overseeing the conversation at a fixed time granularity. The \emph{Director} examines the movements of both interlocutors, analyzes their intentions for the upcoming period, and determines the appropriate interactive behaviors for them. Finally, we develop an translation module that converts the high-level interaction behaviors predicted by the agent into low-level control signals, which then guide the generator in producing interaction behaviors. This creates a continuous feedback loop, enabling dynamic and responsive interactions between the two participants in the conversation.

Our technical contributions can be summarized as: 
\begin{itemize}
    \item We present the first LLM-based agentic framework for generating nonverbal behaviors in dyadic conversations, enabling the synthesis of realistic co-speech body motions with contextually appropriate behaviors across multiple scales. 
    \item We develop a set of knowledge-grounded agentic modules and a control signal space that allow efficient analysis of interlocutors' intentions and the inference of both their actions and reactions in conversations. 
    \item We introduce a dual-person gesture generation model with an interaction guidance strategy, based on auto-regressive diffusion models, enabling high-quality motion synthesis while effectively responding to behavioral control signals.
\end{itemize}

%% file: sec/2_related_work.tex
\section{Related Work}

\subsection{Nonverbal Behavior Generation}
In this paper, we focus on nonverbal behaviors encompassing limited lower-body locomotion, upper-body gestural movements, and head orientation dynamics. Previous studies have primarily focused on single-person co-speech gesture synthesis ~\cite{Nyatsanga_2023}, employing various architectural approaches such as statistical models~\cite{neff2008videogesture}, MLPs~\cite{kucherenko2020gesticulator}, RNNs~\cite{yoon2020trimodalgesture,bhattacharya2021affectivegesture,ghorbani2022zeroeggs,ao2022rhythmicgesticulator}
, CNNs~\cite{habibie2021videogesture,li2021audio2gesture,yi2022talkshow}, Transformers~\cite{qi2023emotiongesturenetease,liu2023emage,Pang_2023,chen2024language}, flow models~\cite{alexanderson2020stylegesture,kucherenko2021speech2properties2gestures}, and diffusion models~\cite{ng2024audio2photoreal,alexanderson2023listendenoiseaction,zhu2023tamingdiffusiongesture,zhang2023diffmotion,zhi2023livelyspeaker,yang2023diffusestylegesture,chen2023diffsheg,yang2024freetalker,liu2024tango} to model behaviors using speech-gesture data. With the emergence of high-quality open-source datasets~\cite{lee2019talking,ghorbani2022zeroeggs,liu2021beatdataset} and the enhancement of stylistic~\cite{ao2023gesturediffuclip} and semantic~\cite{gao2023gesgpt,zhang2024semanticgesture,cheng2024siggesturegeneralizedcospeechgesture} control through large-scale pre-trained models, single-person gesture systems have achieved significant advancements in performance. 
Beyond gestures, head and facial behaviors have also attracted increasing interest. \citet{pan2024s3} incorporate psycho-linguistic insights to design a system that robustly generates 3D head and eye animations for conversing characters, while \citet{ng2022learning2listen,ng2023text2listen} model expressive facial dynamics in dyadic conversations by leveraging semantic and temporal signals from speech.

Recent work has shifted focus toward dyadic interactions, a trend highlighted by the GENEA Challenge 2023~\cite{kucherenko2023geneachallenge2023large}. Many approaches~\cite{mughal2024convofusion,sun2024beyondtalking,huang2024interact,shi2024ittakestwo,qi2025co3gesturecoherentconcurrentcospeech,diffugesture} follow a straightforward paradigm: collecting dual-person audio-gesture data and synthesizing one's behaviors by considering not only their own speech but also the partner's driving signals.
But the complexity of dyadic interactions significantly exceeds that of single-person scenarios, requiring systems capable of high-level behavioral planning, such as interpreting and responding to the partner's emotional states. Simple behavior cloning approaches often fail to generate plausible behaviors. 
\citet{kim2024bodygesture} introduce LLMs for high-level behavioral planning, utilizing the planning outcomes to inform low-level policy in modeling meaningful nonverbal behaviors. They primarily focus on unilateral gestural behaviors of a virtual character interacting with a human user. The difference is that our system models nonverbal behaviors of both participants simultaneously.

\subsection{LLM-based Motion Agent}
Large Language Models (LLMs), with their extensive world knowledge and robust reasoning capabilities, enable high-level semantic guidance for low-level motion generation. 
\citet{liu2024programmableagent} use LLMs to program error functions for open-vocabulary control, \citet{sun2024coma} leverage vision-capable LLMs for motion captioning and trajectory editing, \citet{wu2024motionagent} support interactive generation via function-calling interfaces.
In gesture generation, \citet{torshizi2025largelanguagemodelsvirtual} use LLMs to automate gesture selection, while \citet{windle2024llanimationllamadrivengesture} show LLM-derived text embeddings outperform audio features in beat and semantic gesture synthesis.
In this paper, we leverage LLMs to plan the initial positioning and core joints' trajectories of two characters based on scene context and dialogue content. We then explicitly control the diffusion-based motion policy's outputs through our interaction guidance strategy,  incorporating techniques such as classifier guidance \cite{karunratanakul2023gmd,xie2023omnicontrol}.

\subsection{Nonverbal Behaviors in Human Communication}
Nonverbal behaviors are a cornerstone of human communication and have been extensively studied in psychology and linguistics. One foundational taxonomy by \citet{knapp1972nonverbal} categorizes these behaviors into six major types: Kinesics, Proxemics, Oculesics, Haptics, Facial Expressions, and Paralanguage. Kinesics includes two key types of gesture synchrony. The first is matching, the unconscious imitation of another's gestures, often called the "chameleon effect" \cite{chartrand1999chameleon}. The second is meshing, where a listener provides responsive feedback—such as head nods or thumbs-up gestures—to facilitate mutual understanding. Proxemics concerns the use of interpersonal space and orientation, where physical distance and body arrangement convey significant social meaning \cite{hall1973silent, kendon1990conducting, barua2021detectingsociallyinteractinggroups}. Oculesics involves eye gaze and contact, which are essential for regulating turn-taking, signaling attention, and communicating intent \cite{KENDON196722}. Haptics refers to physical touch between individuals, while Facial Expressions encode emotional and communicative states via micro-expressions. Paralanguage includes vocal elements such as pitch, tone, and volume—excluding the linguistic content itself. In addition to these nonverbal categories, turn-taking coordination constitutes a distinct and essential aspect of human interaction dynamics \citep{SKANTZE2021101178}. 
This work focuses primarily on Kinesics (gesture synchrony), Proxemics (spatial relation), and Oculesics (gaze) as the core modalities for modeling social behavior in our agent system.

%% file: sec/3_approach.tex
\section{Approach}

 \begin{figure*}[t]
  \centering
  \includegraphics[width=0.95\textwidth]{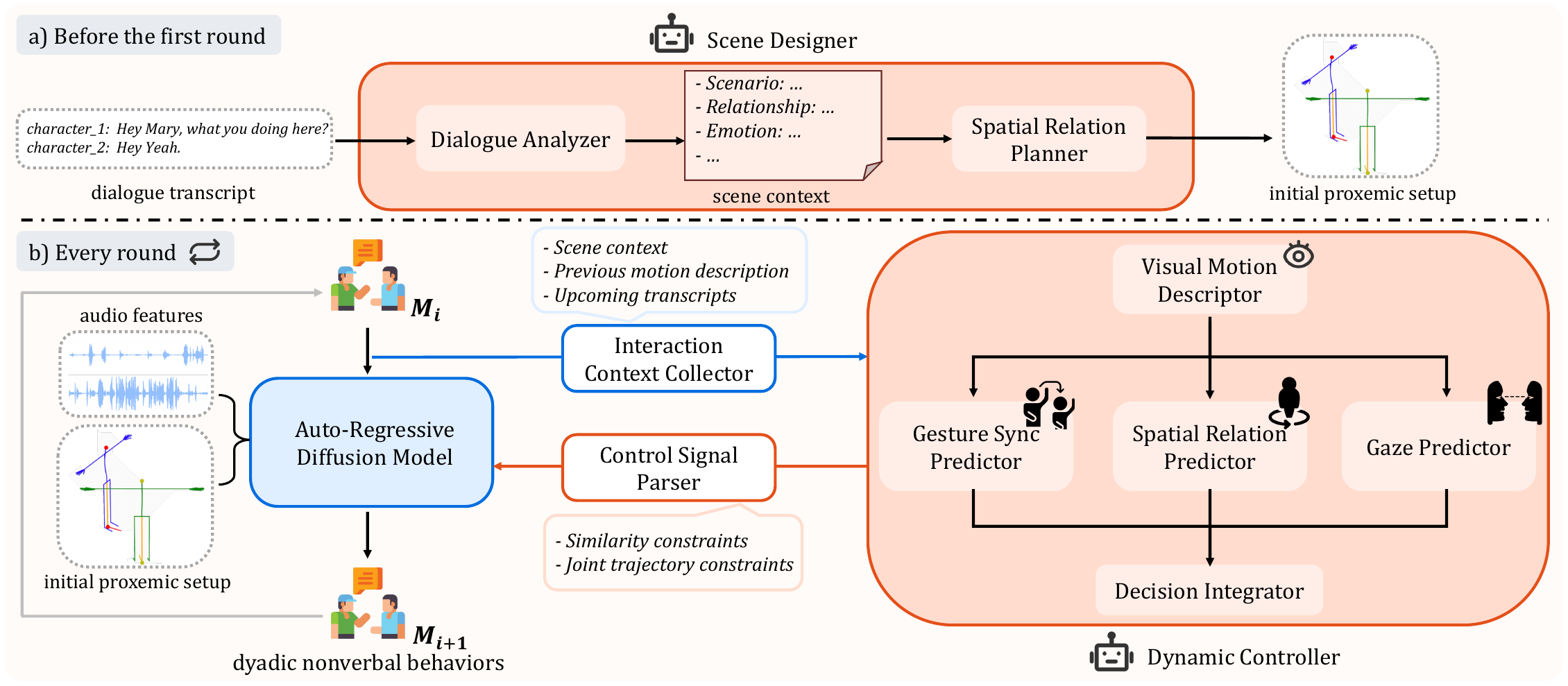}
  \caption{The Social Agent System consists of two key modules: the \emph{Scene Designer}, which analyzes dialogue content to determine the initial proxemic setup at the start of the generation process; and the \emph{Dynamic Controller}, which predicts upcoming interactions for each generation round using multiple predictors. The predicted control signals are then converted into constraints that guide the low-level diffusion model, ensuring coherent and context-aware nonverbal behavior generation.}
  \label{fig:agent}
\end{figure*}

\Cref{fig:overview} illustrates the overall architecture of our system.
Our goal is to synthesize full-body motion sequences for two interlocutors in a dyadic conversation, driven by their audio $(S^{\mathrm{I}}, S^\mathrm{II} )$. The motion sequences, denoted as $(M^{\mathrm{I}},M^{\mathrm{II}})$, each consists of a number of frames $M=[m_t]$, where each frame $m_t \in \mathbb{R}^{(J\times Q + G)}$ encodes both joint-level and global pose information. Here, $J$, $Q$, and $G$ denote the number of joints, joint feature dimension, and global root feature dimension, respectively. 

Our approach consists of three key components. First, we present a dyadic motion generation model (\Cref{sec:dual-person-generative-model}) that effectively synthesizes coordinated dyadic motions from speech inputs. Then \Cref{sec:agent} introduces our LLM-based Social Agent System which can derive contextual interaction constraints between two interlocutors through speech and instruction inputs. Finally, we introduce our training-free motion control mechanism (\Cref{sec:classifier-guidance}) that integrates these interaction constraints to guide the motion generation, significantly enhancing the naturalism and awareness of dyadic nonverbal behaviors.

\subsection{Dual-person Gesture Generative Model}
\label{sec:dual-person-generative-model}

To model the motion distribution of the two interlocutors, $p\left(M^{\mathrm{I}},M^{\mathrm{II}}\right)$, we employ a sliding window mechanism and formulate the problem as a multi-round single-agent motion generation task.
In every round $i$, we generate two motion segments, $(M^{\mathrm{I}}_i,M^{\mathrm{II}}_i)$, for the interlocutors, each consisting of $K$ frames, based on the corresponding chunks of audio $(S^{\mathrm{I}}_i,S^{\mathrm{II}}_i)$. The system then advances to generate the next segment. For character I, this is formalized through the conditional probability distribution:
\begin{equation}
\label{equ:obj_dist}
    p({M}^\mathrm{I}_{i}|{M}^\mathrm{I}_{i-1}, {S}^\mathrm{I}_{i}, {S}^\mathrm{II}_{i}).
\end{equation}
where ${M}^\mathrm{I}_{i}$ is the generation target of Character $I$ in the $i$-th round. ${M}^\mathrm{I}_{i-1}, {S}^\mathrm{I}_{i}$ represent the character's own motion in the previous round and speech features of this round and ${{S}^\mathrm{II}_{i}}$ denote the partner's corresponding features. For Character II, this process is symmetric, with the roles of I and II reversed in the formulation. For simplicity, we proceed with our discussion regarding Character I as the primary agent.

\subsubsection{Full-body Motion Diffusion Model}

We utilize a diffusion model \cite{diffusion_model} to capture the distribution denoted in \Cref{equ:obj_dist}. The training process begins with sampling a clean motion $x_0$, and the forward process follows a Markov chain that gradually adds Gaussian noise to the motion data according to a variance schedule $\beta_t$ $(t=1,...,T)$. At each timestep $t$, the noisy motion $x_t$ is obtained by:
\begin{equation}
    p(x_t|x_{0}) \sim \mathcal{N}(x_t; \sqrt{{ \bar\alpha_t}}x_{0}, (1-\bar\alpha_t)I),
    \label{equ:add_noise}
\end{equation}
where $\bar\alpha_t = \Pi_{s=1}^{t} \beta_{s}$. The reverse process aims to gradually denoise the data by learning to predict the noise component $\epsilon$ at each timestep. Following DDPM \cite{DDPM}, we train the denoiser $\mathcal{D}_{\theta}$ by minimizing the following objective:
\begin{equation}
 \mathcal{L} = \mathrm{E}_{x_0 = {M}_i^{\mathrm{I}},\epsilon \sim \mathcal{N}(0,1),t\in \left[ 0,T\right]}||\epsilon - \mathcal{D}_{\theta}(x_t,t,c)||_2^2,   
\end{equation}
where the conditioning variable $c$ comprises the audio chunks and all possible control signals.

Notably, our model is trained directly in the full-body motion space, unlike prior two-stage approaches~\cite{mughal2024convofusion, sun2024beyondtalking,ao2023gesturediffuclip} that operate in a latent space. This design allows direct control over each joint during generation, enabling motion editing guided by LLM outputs (see \Cref{sec:classifier-guidance}). Compared to latent-space methods, it eliminates the need for backpropagation through a decoder, simplifying constraint enforcement and making joint-level conditioning more flexible and efficient. We also incorporate a condition $s$ within $c$ to regulate the motion state of the character, which can be one of three possible states: $s \in \{\text{stand}, \text{walk}, \text{sit} \}$.

We utilize the classifier-free guidance (CFG)~\cite{ho2022classifierfreediffusionguidance} mechanism  to enhance the model's compliance with speech inputs. Specifically, during the training phase, we randomly set ${S}^{\mathrm{I}}_{i} = \varnothing$ or ${S}^{\mathrm{II}}_{i} = \varnothing$ with a probability of $p$. During the inference phase, the predicted noise is computed using:
\begin{equation}
    \begin{aligned}
        \mathcal{D}_{\theta}(x_t, t, c) = 
        & \, \lambda \mathcal{D}_{\theta}\left(x_t, t, s; {M}^\mathrm{I}_{i-1}, {S}^\mathrm{I}_{i}, {S}^\mathrm{II}_{i} \right) \\
        & + (1-\lambda) \mathcal{D}_{\theta}\left(x_t, t, s; {M}^\mathrm{I}_{i-1}, \varnothing, \varnothing \right).
    \end{aligned}
\end{equation}
This scheme allows us to control the effectiveness of the speech input with the scale factor $\lambda$. Details of our model architecture are provided in \Cref{appendix_model_arch}.

\subsection{LLM-based Social Agent System}
\label{sec:agent}

Our approach leverages an LLM-based agentic system, to derive contextual interaction constraints for nonverbal behavior generation in dyadic conversation scenarios. This system is designed to act as a \emph{Director} and provide high-level guidance for nonverbal behavior by analyzing multimodal inputs and instruction prompts. As shown in \Cref{fig:agent}, the system comprises two main components: the \emph{Scene Designer Agent}, which operates before the initial round to analyze the dialogue and determine the initial proxemic setup, and the \emph{Dynamic Controller Agent}, which is activated at the beginning of each round to analyze the current state, interpret the intentions of the interlocutors and determine the appropriate interactive behaviors for them. All modules in the Agent system are built into the prompt design method, using carefully tailored prompts based on relevant linguistic and human behavioral research.

\subsubsection{Scene Designer Agent}
\label{sec:sda_32}

As illustrated in \Cref{fig:agent}a, given the audio of a dynamic conversation as input, we perform automatic speech recognition \cite{radford2022robustspeechrecognitionlargescale} on the audio to obtain dialogue transcripts. These transcripts are then processed by the \emph{Dialogue Analyzer}, which extracts relevant scene context, such as the scenario, relationships between the participants, emotion, and character settings. 
The \emph{Spatial Relation Planner} then analyzes this context to construct the initial spatial layout of the scene—determining each character’s postural state, global position, and orientation. Due to the limitations of current LLMs in direct spatial reasoning, we design a structured prompting and reasoning process: instead of directly predicting 3D coordinates, the agent first infers high-level qualitative spatial relationships between interlocutors, which are later translated into quantitative values. Specifically, the agent performs chain-of-thought reasoning to generate three core aspects of proxemic configuration:

\begin{itemize}
    \item \emph{Positional Configuration}. According to Kendon’s F-formation system \cite{kendon1990conducting, barua2021detectingsociallyinteractinggroups}, the spatial arrangement between two characters can be categorized as Vis-à-vis, L-shaped, or side-by-side, based on conversational context.
    \item \emph{Spatial Distance}. Based on Hall’s proxemics theory \cite{hall1973silent}, interpersonal distance can be categorized into Interpersonal, Social, or Public categories.
    \item \emph{Postural State}. Whether a character is sitting or standing. 
\end{itemize}

The LLM agent then translated these qualitative outputs into numerical spatial parameters using predefined mapping rules. For instance, a positional configuration like vis-à-vis is first mapped to two directional relationships (e.g., “A is in front of B”), which are then converted into clock-based angles (e.g., “A is at B’s 11:50 direction”). Distance categories are similarly mapped to fixed metric ranges. These mapping rules are provided to the LLM through structured prompts. Finally, by anchoring Character I at a fixed origin, we compute Character II’s global position and orientation based on the predicted relative values, establishing the initial proxemic setup for motion generation.

\subsubsection{Dynamic Controller Agent}
As shown in \Cref{fig:agent}b, the \emph{Dynamic Controller} Agent is called at the beginning of every round to analyze the current state and then output interaction adjustment signals for the upcoming round. 

The input to this module is gathered by the \emph{Interaction Context Collector}, which contains multimodal information including: a) scene context from the Dialogue Analyzer, b) descriptions of the previous motion, detailing the relative orientation and distance between the two characters, as well as the directions of their head orientation from the last generated frame of the previous round, and c) upcoming dialogue transcripts for the next round. This information is converted into natural language using a set of templates. Additionally, We employ a vision-language module as the \emph{Visual Motion Descriptor} to generate a description of the movements of the interlocutors, particularly focusing on upper body gestures, using a rendered image of their current poses. This approach provides the agent with a richer, multi-faceted understanding of the scene, enabling it to generate more contextually appropriate interactions.

This comprehensive interaction contexts are then processed by three interactive processing channels at different behavioral scales, each dedicated to a different aspect of nonverbal behavior in dyadic interactions. We prompt these modules with findings from the literature in psychology and linguistics, allowing them to leverage established knowledge for more informed analysis.

\emph{Spatial Relation Predictor} assesses whether adjustments in position and orientation will occur in the next round. Similar to the Spatial Relation Planner, this module first determines whether Positional Configuration changes are required for each character individually. The updated orientation around the vertical axis is then computed using the predefined mapping rules. Additionally, it predicts whether the characters will move closer or farther apart and estimates the target constraints for their global horizontal positions.

\emph{Gesture Sync Predictor} models two types of gesture synchrony in interaction: matching and meshing \cite{knapp1972nonverbal}. 
The module analyzes the current scene context to predict whether synchrony will occur and which type is most likely. It also identifies the roles of each participant: who will initiate the gesture and who will respond, either through imitation or nodding. To further pinpoint the timing of the imitation, the module also predicts which word in the transcript corresponding timestamp of this word is then extracted as the gesture synchrony timestamp.

\emph{Gaze Predictor} forecasts whether one character will look at the other in the next round. 
The agent analyzes the current scene to determine whether mutual gaze will occur and estimates its duration. Like the Gesture Sync Predictor, this module identifies the specific word in the transcript most likely to trigger gaze and retrieves the corresponding timestamp, establishing the timing of the gaze event.

After the three prediction modules propose their suggestions, the \emph{Decision Integrator} aggregates and integrates them into a cohesive adjustment suggestion. Based on the current scene context, it individually selects the most appropriate adjustment combination for each of the two characters from the three proposals or determines that no adjustment is necessary. The natural language descriptions of the adjustments are finally converted into digital control signals by the \emph{Control Signal Parser (\Cref{sec:classifier-guidance})}, and then fed back to the generative model to guide the next round of interaction.

\begin{figure*}[htp]
  \centering
  \includegraphics[width=\textwidth]{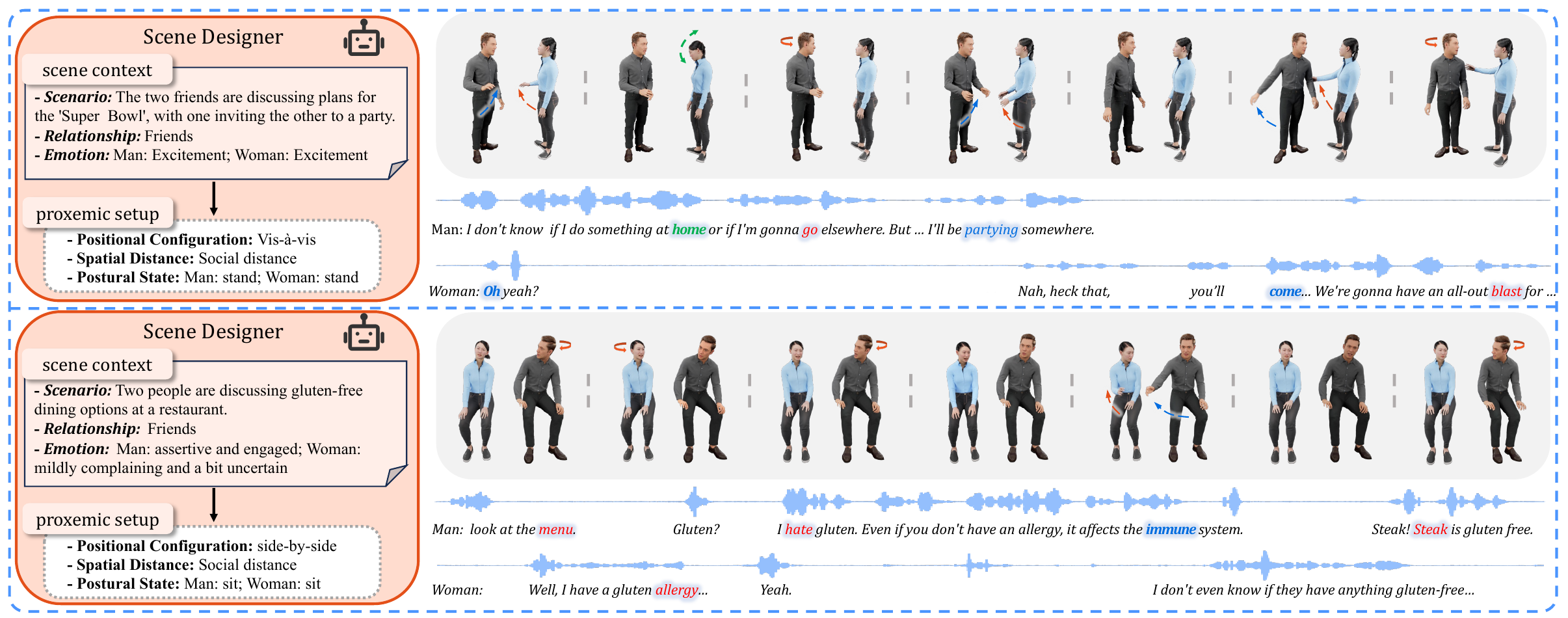}
  \caption{Dyadic nonverbal behaviors generated by our system. Left: Scene Designer predicts the initial proxemic setup. Right: Dynamic Controller’s signals with corresponding target word (red for gaze, blue for gesture imitation, and green for nodding). Motion trend line show imitation patterns (blue: imitated character, red: imitator). The Scene Designer ensures scene-aware spatial arrangements, while the Dynamic Controller guides cohesive dyadic interactions.}
  \label{fig:res}
\end{figure*}

\subsection{Interaction Guided Motion Generation}
\label{sec:classifier-guidance}

Building upon the pre-trained diffusion model’s capability to generate realistic gestures, we introduce an interaction guidance strategy to enforce adherence to interaction constraints specified by the LLM agent. 
More precisely, our framework employs a \emph{Control Signal Parser} that processes the structured JSON output from the LLM agent to extract motion adjustment signals. These signals are translated into motion constraints via predefined rules, and categorized into two types: similarity constraints and joint trajectory constraints. Specifically, gesture imitation signals are converted into similarity constraints, while numerical adjustments in position and orientation are transformed to root trajectory constraints. Nodding cues are interpreted as head trajectory constraints, simulated by applying a sinusoidal function to the head’s pitch angle.
Gaze signals are handled by computing the head orientation required to face the partner’s head, which is then encoded as a head trajectory constraint.

For similarity constraints, although previous works have explored keyframe-based motion editing methods \cite{diffusionprior}, our goal is not to achieve exact motion correspondence but rather to maintain general similarity. Therefore, we adopt a straightforward yet effective approach: replacing the motion with the target motion $\tilde{x}$ during the early stages of the denoising process:
    ${x}_{t<\tilde{t}}^0 = \tilde{x}$,
where $\tilde{t}$ is a predefined step.

For the trajectory constraints, we transform them into a mathematical formulation through a loss function $\mathcal{L}(x_t^0) = \| W \odot (J(x_t^0) - \tilde{J}) \|$, where $x_t^0$ is the predicted clean motion at denoising step $t$, and $J(\cdot)$ represents the extraction operator that maps $x_t^0$ to its corresponding joint parameters, and $\tilde{J}$ denotes the target joint trajectory. And $W$ is a mask matrix with the same dimensions as  $J(x_t^0)$  and  $\tilde{J}$ , containing 1 for joints that should be constrained and 0 for those that should be ignored. Following \cite{karunratanakul2023gmd,xie2023omnicontrol}, we use the gradient of $\mathcal{L}(x_t^0)$ to guide the denoised motion at each denoising step with a guidance strength factor $\alpha$:
\begin{equation}
    \tilde{x}_t^0 = x_t^0 - \alpha \nabla_{x_t^0} \mathcal{L}(x_t^0)
    \label{equ:cguidance}
\end{equation}
To enhance guidance effectiveness while maintaining motion quality, we apply two gradient updates per step during the first  $\tau$  portion of the denoising steps, where  $\tau$  defines the control scope. Additionally, for root trajectory constraints, we set the next round’s state $s =  walk$ to ensure natural and coherent leg movements during root adjustments.

%% file: sec/4_experiment.tex
\section{Experiment}

\begin{figure*}[htp]
  \centering
  \includegraphics[width=0.8\textwidth]{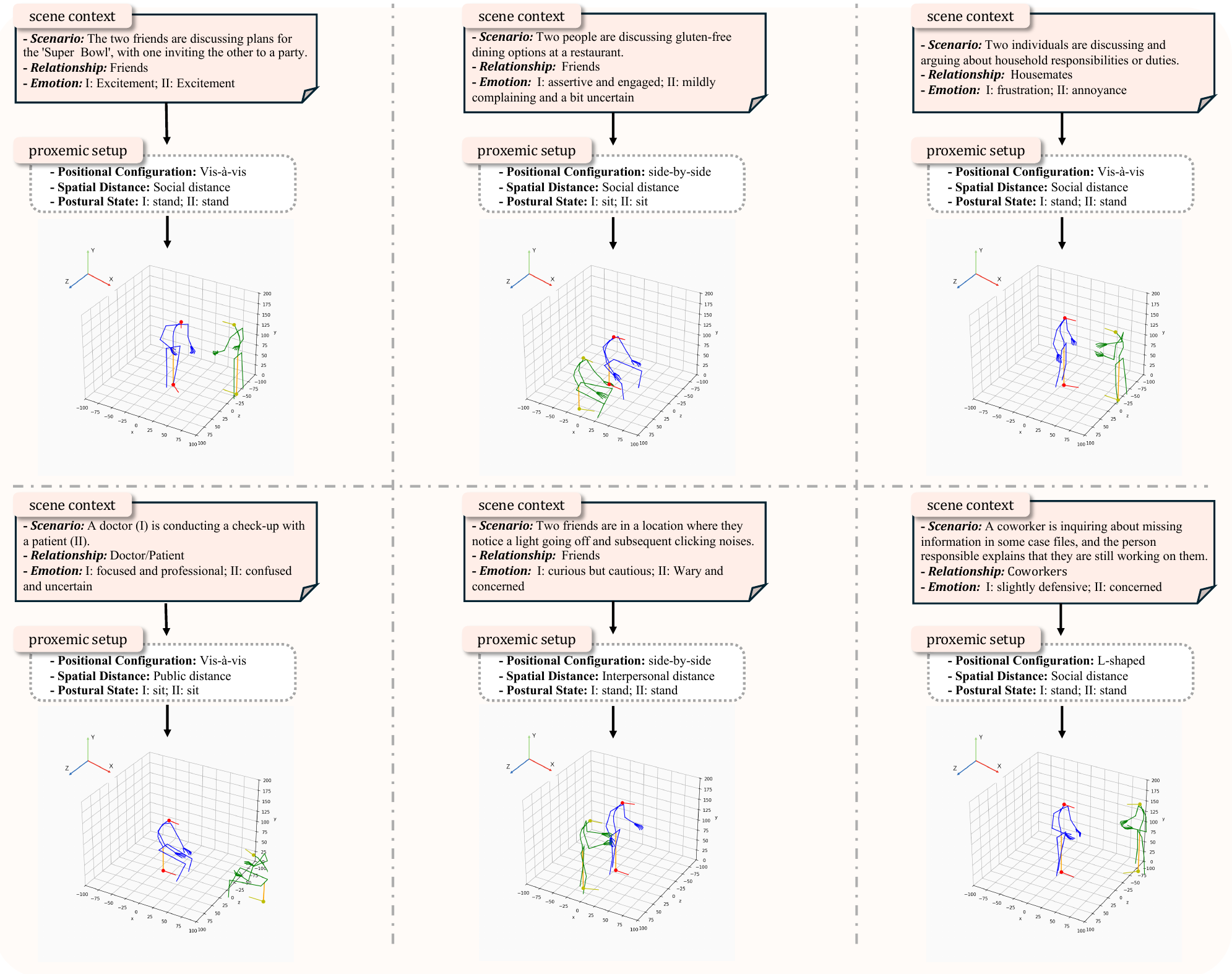}
  \caption{Visualization of the Scene Designer Agent process workflow and results. The blue character is Character I, and the green character is Character II. The examples showcase the framework's scene analysis and understanding capabilities, illustrating how it designs realistic and contextually appropriate initial proxemic setups for different scenarios. This facilitates subsequent interaction control by the Dynamic Controller Agent module, ensuring more natural and context-aware interactions.}
  \label{fig:sda_res}
\end{figure*}

\subsection{System Setup}
\subsubsection{Speech-Gesture Datasets}

We base our experimental evaluation on two high-quality, publicly available speech-gesture datasets: the Photoreal dataset \cite{ng2024audio2photoreal} and the InterAct dataset \cite{huang2024interact}.  
The InterAct dataset provides both motion and audio for both participants, while the Photoreal dataset contains motion for only one speaker and audio for both. 
Specifically, to ensure compatibility with our model, we converted the parametric motion format in Photoreal dataset into BVH skeletal data using the official code.
Comprehensive descriptions of the datasets and our preprocessing procedures are provided in Appendix~\ref{appendix_data_process}.

\subsubsection{Settings}
For the $s \in \{\text{stand}, \text{walk}, \text{sit} \}$, we use three learnable embeddings corresponding to the three distinct states. Our model consists of $6$ blocks, with $8$ attention heads in the attention layer and a hidden state dimension of $1280$. During training, we set diffusion step $T=1000$, window size $K=150$, and apply a dropout probability of $p=0.2$ to the condition. For the Social Agent system, we employ gpt-4o-2024-08-06 \cite{openai2024gpt4o} as the LLM model and construct precise prompts tailored for it. The detailed prompt examples are provided in the Supplementary Materials. During the inference phase, we employ a 200-step DDIM \cite{DDIM} acceleration. We set the classifier-free guidance scale factor $\lambda=2$, the similarity constraint replacement cutoff step $\tilde{t} = 200$, and the control scope $\tau = 80\%$. Regarding classifier guidance, our setup follows a similar approach to \citet{xie2023omnicontrol}. To refine motion control, we adapt the guidance strength based on the variance of joint motion, applying $\alpha=\{0.1,20,100 \}$ respectively to the root displacement, root rotation, and head rotation. We train for $300$ epochs on both datasets, using a learning rate of $10^{-4}$. 
The training process takes approximately $8$ hours on four state-of-the-art consumer GPUs.

\begin{figure}[htb]
  \centering
  \includegraphics[width=0.95\columnwidth]{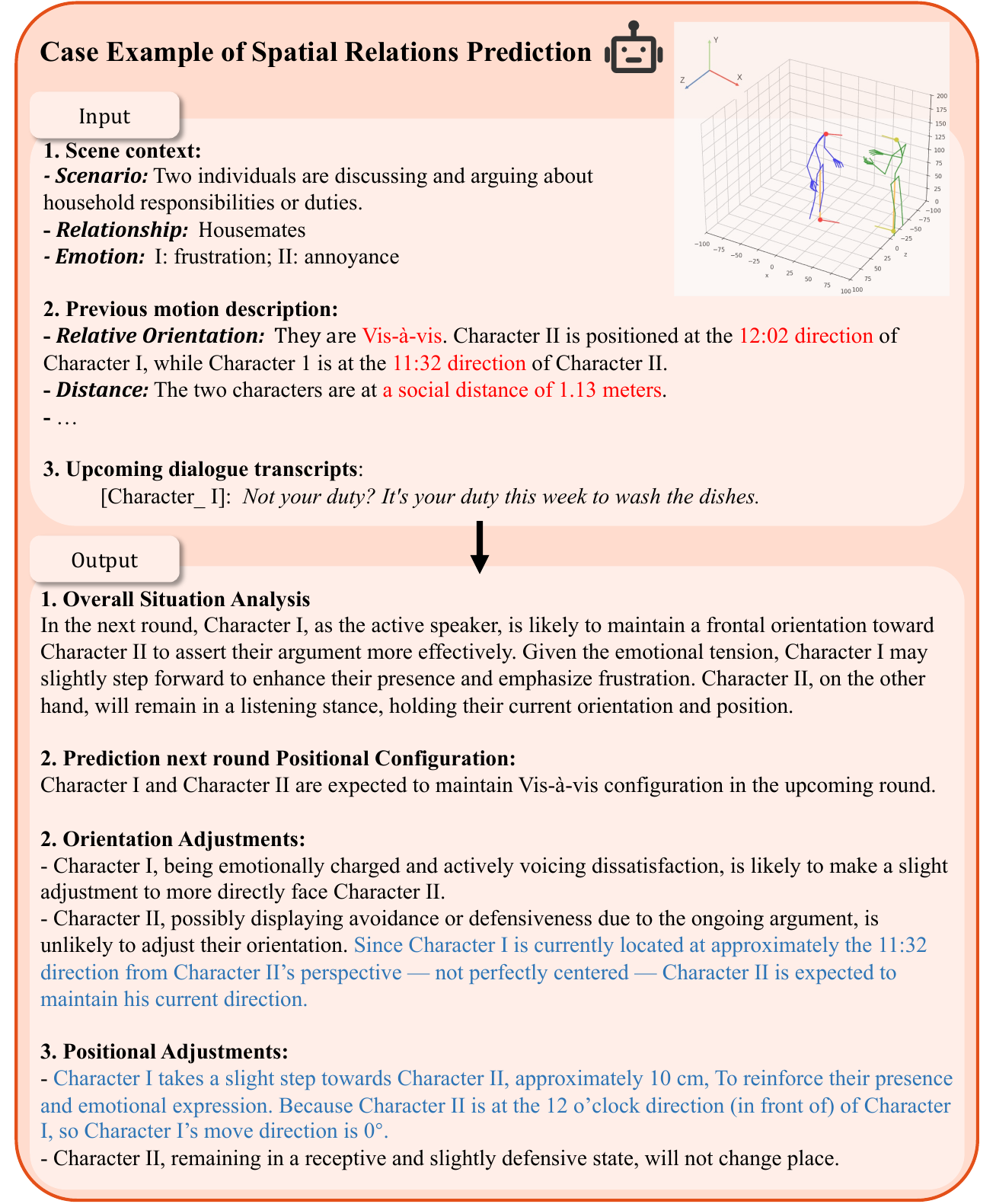}
  \caption{
  This example illustrates how the \emph{Spatial Relation Predictor} conducts fine-grained spatial reasoning based solely on textual input. Red text in the input highlights the current spatial state of both characters. The 3D image on the right visualizes the input configuration but is not part of the model’s input. In the output, blue text emphasizes the model’s spatial reasoning process, such as the inferred direction and distance of Character I’s movement. This is a concise version of the agent’s output, preserving essential information.
  }
  \label{fig:example}
\end{figure}

\subsection{Results}

As illustrated in \Cref{fig:res}, our system generates dyadic nonverbal behavior based on several in-the-wild audio pairs and interaction control signals from the Social Agent system. We employ the MetaHuman plugin of Unreal Engine \cite{unreal_audio2face} to produce facial animations from audio. The results demonstrate that our system successfully synthesizes high-quality, realistic dyadic interactions, enhancing the naturalness and coherence of dialogue scenarios. On the left side, we showcase the Scene Designer workflow, which extracts scene context and generates the initial proxemic setup. This module proves effective in analyzing and structuring the initial interaction scene. For instance, when two friends are ordering food in a restaurant, the system positions them sitting side by side. Additional results of this module can be found in \Cref{fig:sda_res}. On the right side, the generated motion sequences demonstrate that the Dynamic Controller module effectively captures the interaction intentions and produces multiscale precise interaction signals, such as gaze, nodding, and gesture imitation. These high-level signals further guide the generative model to synthesize realistic and coherent interactions.
Moreover, \Cref{fig:example} illustrates the Dynamic Controller Agent’s capability for complex spatial reasoning, enabling it to interpret textual input to generate fine-grained spatial predictions.

\subsection{Comparison}

\begin{figure}[!htb]
  \centering
  \includegraphics[width=\columnwidth]{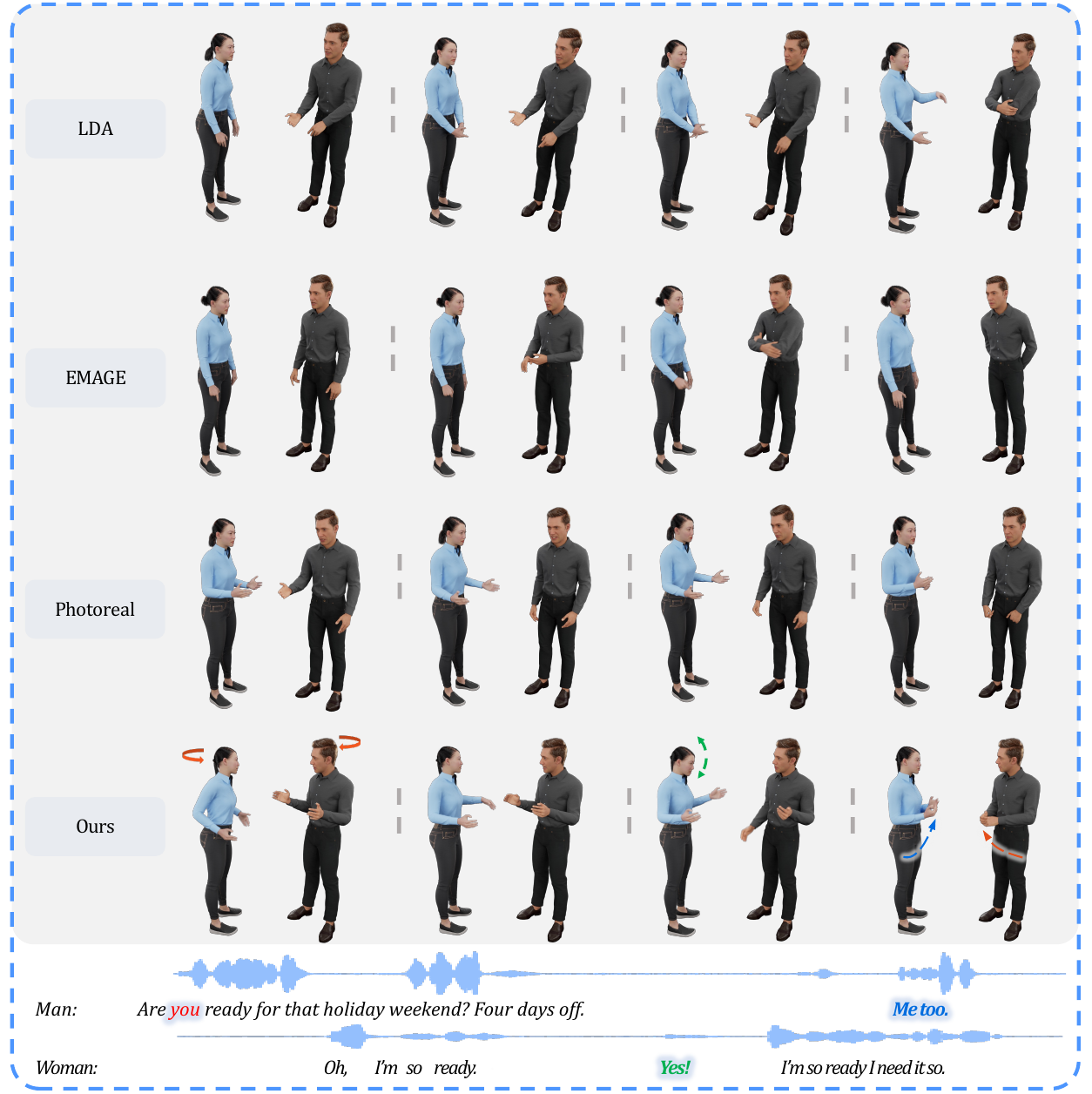}
  \caption{Qualitative comparisons: Ours vs. LDA~\cite{alexanderson2023listendenoiseaction}, EMAGE~\cite{liu2023emage}, and Photoreal~\cite{ng2024audio2photoreal} on the Photoreal dataset.}
  \label{fig:com_photoreal}
\end{figure}
\begin{figure}[!htb]
  \centering
  \includegraphics[width=\columnwidth]{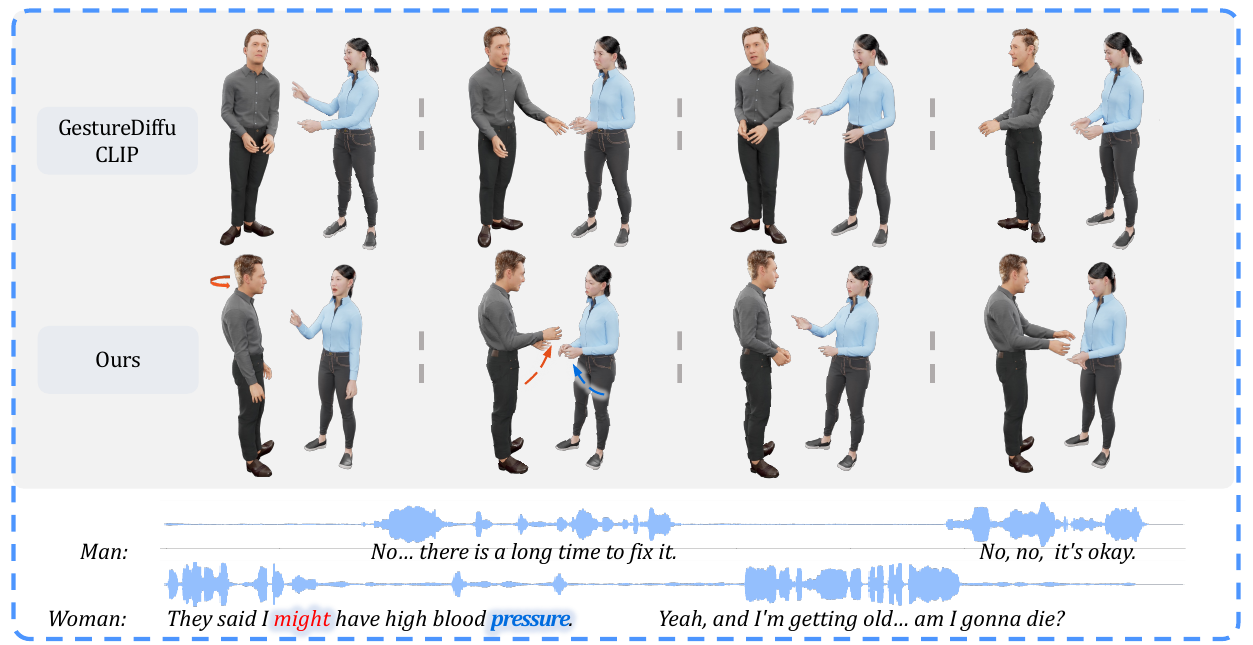}
  \caption{Qualitative comparisons: Ours vs. GestureDiffuCLIP~\cite{ao2023gesturediffuclip} on the InterAct dataset.}
  \label{fig:com_interact}
\end{figure}

Evaluating non-verbal behaviors (e.g., gestures) using objective metrics presents numerous challenges, as many existing objective metrics have a low correlation with subjective feedback \cite{Kucherenko_2024}. In line with the approaches outlined in \cite{alexanderson2023listendenoiseaction,ao2023gesturediffuclip,zhang2024semanticgesture}, this study relies on user evaluations to assess the generated results, with quantitative evaluation serving as an auxiliary reference.

\subsubsection{Baselines}

At the time of writing, the source code for existing dyadic gesture generation systems \cite{shi2024ittakestwo,huang2024interact,sun2024beyondtalking,qi2025co3gesturecoherentconcurrentcospeech,diffugesture} remains unavailable. Additionally, while some methods claim to support dyadic conversational scenarios, they can only generate gestures for a single individual at a time \cite{mughal2024convofusion,ng2024audio2photoreal,kim2024bodygesture}, making them unsuitable for modeling interactive behaviors between two individuals. Due to this limitation, no suitable dyadic systems are available for direct comparison. 
We instead compare against state-of-the-art single-person gesture generation models: LDA \cite{alexanderson2023listendenoiseaction}, EMAGE \cite{liu2023emage}, and Photoreal \cite{ng2024audio2photoreal} on the Photoreal dataset, and GestureDiffuCLIP \cite{ao2023gesturediffuclip} on the InterAct dataset. We re-train LDA, EMAGE, and GestureDiffuCLIP on the corresponding datasets, and use the publicly released checkpoint for Photoreal.
To simulate dyadic motions using single-person models, we perform separate inferences on each audio stream within the dyadic audio pair. For fair comparison, we align both the initial positions and orientations of the generated characters with our model’s output. Motions generated by Photoreal model are converted to skeletal format for unified evaluation.

\begin{table}[t]
    \centering
    \caption{Average scores of user study with 95\% confidence intervals. Ours (w/o DCA) excludes the Dynamic Controller Agent for the pre-trained generator. Asterisks indicate the significant effects.}
    \resizebox{\columnwidth}{!}{
        \begin{tabular}{llccc}
            \toprule
            Dataset & System & Human Likeness $\uparrow$ & Beat Matching $\uparrow$ & Interaction Level $\uparrow$  \\ 
            \toprule
            \multirow{5}*{Photoreal} & LDA  & -0.20$^*$  & -0.08$^*$ & -0.16$^*$ \\
            & EMAGE  & -0.25$^*$  & -0.04$^*$ & -0.15$^*$  \\
            & Photoreal  & 0.10$^*$ & 0.03 & -0.07$^*$   \\
            & Ours (w/o DCA)  & 0.09$^*$  & 0.04 &  0.02$^*$ \\
            & Ours  & \textbf{0.26}  & \textbf{0.04} & \textbf{0.37}    \\
            
            \midrule
            \multirow{4}*{InterAct} & GT    & 0.42$^*$  & 0.14$^*$ & 0.38$^*$  \\ \cmidrule(lr){2-5}
            & GestureDiffuCLIP    & -0.31$^*$  & -0.05 & -0.26$^*$ \\
            & Ours (w/o DCA)    & -0.19$^*$  & -0.03 & -0.16$^*$ \\
            & Ours    & \textbf{0.08}  & -0.03 & \textbf{0.11}  \\
            \bottomrule
        \end{tabular}
    }
    \label{tab:user_study}

\end{table}

\subsubsection{User Study}
\label{subsubsec_user_study}

Following the approach outlined in \cite{alexanderson2023listendenoiseaction,ao2023gesturediffuclip,zhang2024semanticgesture}, we conduct user studies through pairwise comparisons, recruiting participants via the Credamo platform \cite{credamo}. Three distinct preference tests are carried out: \emph{human likeness}, \emph{beat matching}, and \emph{interaction level}. A detailed description of the user study as well as the definitions of these evaluation metrics is provided in \Cref{appendix_user_study}.

On the Photoreal dataset, we compare five methods: our full model (Ours), an ablated version without the Dynamic Controller Agent (w/o DCA), LDA, EMAGE, and Photoreal. On the InterAct dataset, we compare four methods: ground truth (GT), Ours, Our (w/o DCA), and GestureDiffuCLIP. As shown in \Cref{tab:user_study}, while \emph{beat matching} results are comparable across methods, our model significantly outperforms baselines in terms of \emph{human likeness} and \emph{interaction level}, underscoring the importance of the Social Agent System. \Cref{fig:com_photoreal} and \Cref{fig:com_interact} further demonstrate the improved interactive quality of the generated motions across both datasets. Compared to other methods, our results show more natural and synchronized nonverbal behaviors, indicating stronger engagement between the two individuals.

\subsubsection{Quantitative Evaluation}
We quantitatively evaluate the motion quality and interaction realism using a composition of metrics: a) Fr{\'e}chet Gesture Distance (FGD) \cite{yoon2020trimodalgesture} quantifies the disparity between the latent feature distributions of generated and real gestures, evaluating gesture perceptual quality; b) BeatAlign \cite{li2021aichoreographermusicconditioned} assesses speech-motion synchrony by measuring the temporal alignment between motion beat candidates; c) Fr{\'e}chet Distance-Matrix Distance (FDD) \cite{shi2024ittakestwo} quantifies the disparity between the per-joint distance matrices of generated and real interactive motion pairs using the Fr{\'e}chet Distance, measuring interaction realism. 

To further assess the temporal consistency of interaction dynamics, we introduce the Delayed Motion Synchrony Score (DMSS), inspired by cognitive psychology studies on global synchrony \cite{ng2022learning2listen, boker2002windowed}. Unlike FDD, which focuses on spatial interaction fidelity, DMSS captures dynamic coupling over time, accounting for phase-shifted behaviors such as turn-taking or responsive gestures. It computes the maximum Pearson correlation between the joint velocity sequences of two individuals over a range of temporal lags, allowing for flexible alignment. A higher DMSS indicates stronger motion coordination. Full computational details are provided in Appendix \ref{appendix_dmss}.

As shown in Table~\ref{tab:Quantitative}, our system outperforms all baselines on BeatAlign, FDD, and DMSS across both the Photoreal and InterAct datasets. For FGD, our model performs competitively—slightly below the Photoreal upper bound trained on in-domain ground-truth data, yet significantly surpassing all other baselines. In particular, our method achieves notable improvements on FDD and DMSS, indicating more realistic, temporally coordinated, and socially responsive interactive motions. These results validate the effectiveness of our framework in generating high-quality, socially coordinated dyadic nonverbal behaviors.

\subsection{Ablation Study}
\label{ablation}

\begin{table}[t]
    \centering
    \caption{Quantitative evaluation on the Photoreal and InterAct datasets. All methods are trained on the same training data, and evaluated on the test audio. Note that FDD cannot be computed on the Photoreal dataset, as it lacks ground-truth paired two-person motion sequences.}
    \resizebox{0.9\columnwidth}{!}{
        \begin{tabular}{llcccc}
            \toprule
            Dataset & System & FGD$\downarrow$ & BeatAlign $\uparrow$ & FDD $\downarrow$ & DMSS $\uparrow$ \\ 
            \toprule
            \multirow{5}*{Photoreal} & LDA & 78.67  &  0.736 & -  & 0.235 \\
            & EMAGE  & 83.58  & 0.764 & -  & 0.247 \\
            & Photoreal  & \textbf{68.93}  & 0.751 & -  & 0.279 \\
            & Ours (w/o DCA)  & 73.31  & 0.818 & -  & 0.254 \\
            & Ours  & 71.22  & \textbf{0.827} & -  & \textbf{0.457}  \\
            \midrule
            \multirow{3}*{InterAct} & GestureDiffuCLIP  & 107.88   & 0.759 & 143.12  & 0.216 \\
            & Ours (w/o DCA)  & 95.13  & 0.794  & 120.39  & 0.237 \\
            & Ours  & \textbf{90.48}  & \textbf{0.802}  & \textbf{105.16}  & \textbf{0.439} \\
            \bottomrule
        \end{tabular}
    }
    \label{tab:Quantitative}
\end{table}

\subsubsection{Prompt of Social Agent System.}
This experiment evaluates the impact of prompt quality on the reasoning ability of the LLM agent. Our deliberate designed prompt includes: reference behavioral theories, mapping rules particularly for spatial relations, task definition, and stepwise reasoning guides which means explicit decomposition of reasoning steps. We compare three prompt variants: a) Baseline—containing only mapping rules and task definition; b) + Stepwise Reasoning Guides; and c) + Reference Theories (our complete prompt).
We evaluate 20 scenes each for the Scene Designer and Dynamic Controller Agents using the same LLM. For Scene Designer, we assess \emph{Layout Plausibility}; for Dynamic Controller, \emph{Behavior Appropriateness}. Outputs are judged by GPT-4.1~\cite{openai2025gpt41} following the LLM-as-a-judge protocol~\cite{zhang2023wider}, using a 1–5 rating scale. See \Cref{appendix_prompt_ablation} for more details.

\begin{figure}[tb]
  \centering
  \includegraphics[width=\columnwidth]{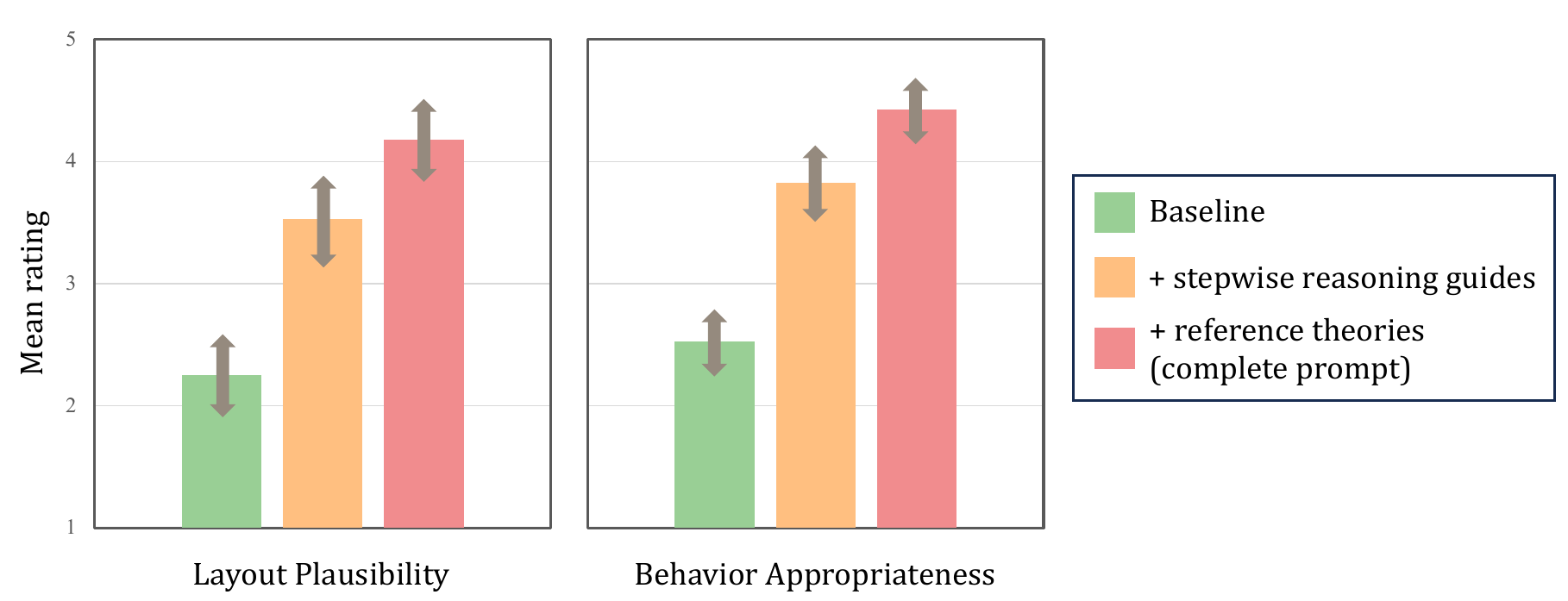}
  \caption{Ablation study on prompt design. We evaluate three prompt variants of increasing complexity: a) Baseline—containing only mapping rules and task definition; b) + Stepwise Reasoning Guides; and c) + Reference Theories (complete prompt). Each configuration is assessed on two metrics: Layout Plausibility (for the Scene Designer Agent) and Behavior Appropriateness (for the Dynamic Controller Agent). Ratings are given by GPT-4.1 on a 1–5 scale. Results show that adding stepwise reasoning substantially boosts performance, and incorporating reference theories further improves outcomes, underscoring the effectiveness of structured prompting.}
  \label{fig:com_ablation}
\end{figure}

The final results are shown in \Cref{fig:com_ablation}. As illustrated, adding stepwise reasoning leads to a substantial improvement in performance across both agents and evaluation metrics. Incorporating reference theories on top of stepwise reasoning provides an additional performance gain, indicating that both components contribute positively. These findings highlight the critical role of structured prompting in improving the reasoning quality of LLM agents.

\subsubsection{Architecture of Social Agent System.}

\begin{figure}[tb]
  \centering
  \includegraphics[width=\columnwidth]{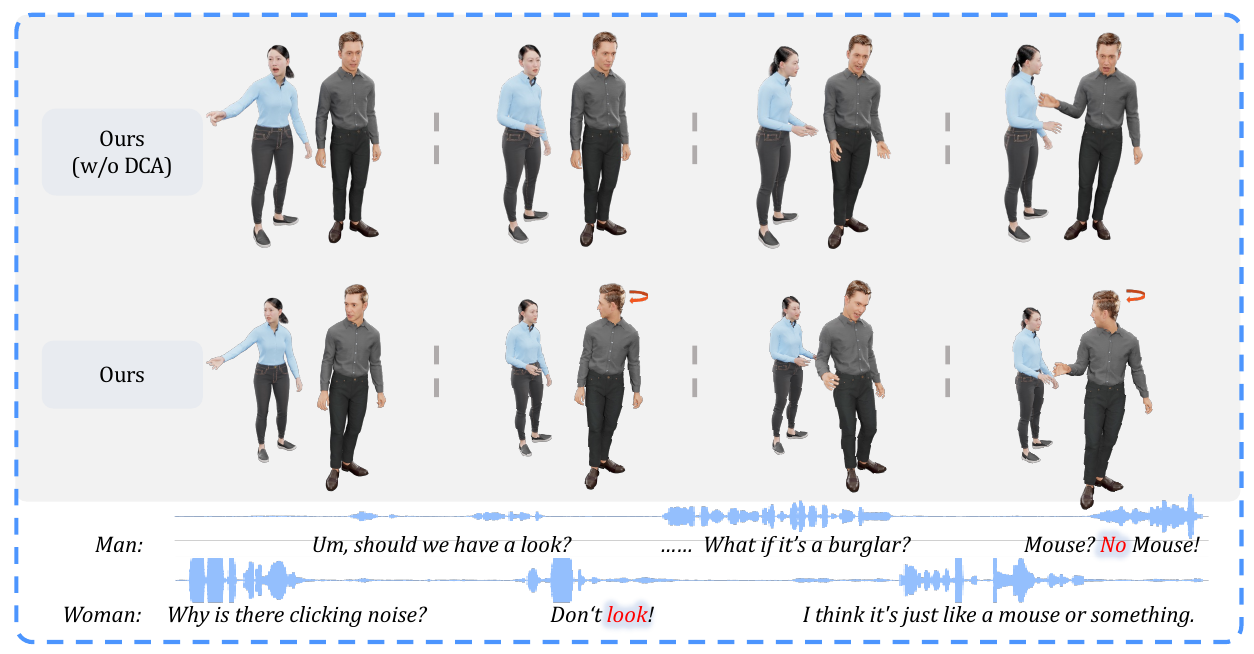}
  \caption{Qualitative comparisons: Ours vs. our ablated model without the Dynamic Controller Agent (w/o DCA).}
  \label{fig:ablation_1}
\end{figure}

This experiment evaluates the importance of the Dynamic Controller module. Specifically, we remove this module and conduct experiments, leading to a notable decline in interaction-related metrics, such as Interaction Level (\Cref{tab:user_study}), FDD and DMSS (\Cref{tab:Quantitative}). As illustrated in \Cref{fig:ablation_1}, the absence of this module results in the model losing high-level guidance during motion generation, causing a lack of awareness of interaction cues in the generated motions. 

\subsubsection{Interaction Guidance Strategy.} In this experiment, we explore the effect of the control scope parameter $\tau$  for classifier guidance by testing three different values: 100\%, 80\%, and 50\%. Our findings indicate that: A smaller control scope results in insufficient guidance, while a larger control scope degrades motion quality, introducing instabilities and jitter. For additional visualization results, please refer to the supplementary video.

%% file: sec/5_conclusion.tex
\section{Conclusion}

In this paper, we introduce Social Agent, a framework for dyadic nonverbal behavior generation in conversations. We first develop a diffusion-based model for auto-regressive dyadic gesture generation. Building upon this, we design an interaction-aware agentic framework that analyzes scene context and generates interaction control signals. Finally, an interaction guidance strategy translates these signals into corresponding interactive motions. Visualization results show that our system produces high-quality and realistic dyadic nonverbal behaviors. Furthermore, user studies and quantitative evaluations confirm the superiority of our framework. 

Despite its effectiveness, our approach has several limitations that offer directions for future work. 
First, our system can generate gaze behavior at a higher frequency, which is desirable in scenarios such as television interviews but may appear less natural in other contexts. This can be addressed by applying the system to more diverse character types with corresponding contexts, or by adjusting the LLM prompts for different interaction settings. 
Second, a potential concern is the unnaturalness of certain nodding behaviors. This issue stems from their scarcity in the training data, which required procedural generation under strong constraints. Incorporating datasets with richer feedback behaviors would help address this limitation. Additionally, motion artifacts such as foot-sliding remain to be resolved through post-processing techniques. 
Finally, our current behavior set focuses on the most common interaction types. Future extensions may involve modeling more complex nonverbal behaviors (e.g., physical contact) and holistic generation with eye contact to enhance expressiveness.

%% file: sec/appendix.tex
\clearpage

\appendix

\section{Model Architecture}
\label{appendix_model_arch}

\begin{figure}[tb]
  \centering
  \includegraphics[width=\columnwidth]{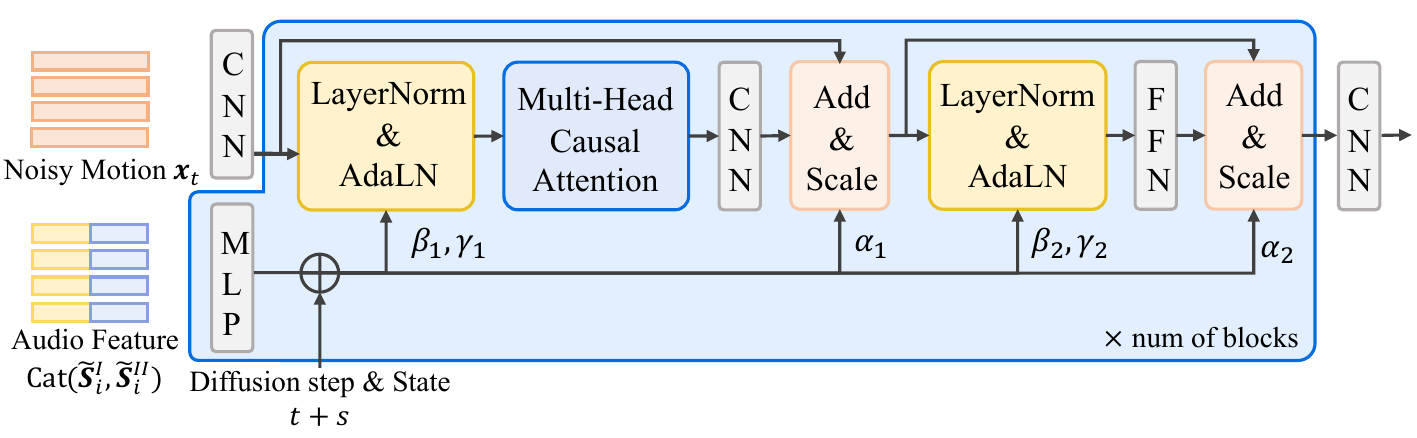}
  
  \caption{Model Architecture. Our network architecture is based on the Transformer \cite{Attentionisallyouneed} block, augmented with a CNN module to enhance the learning of local temporal features. Additionally, the audio conditions $\tilde{S}^\mathrm{I}, \tilde{S}^{\mathrm{II}}$, as well as the speaker’s motion state $s$, are transformed via an MLP into several scale factors. These factors are used to scale the features at each layer, thereby guiding the generation process. }
  \label{fig:architecture}
\end{figure}

Our model architecture is shown in \Cref{fig:architecture}. For the input noisy action $x_t$, we first use a temporal CNN network to extract its features. Then, the features are processed through several identical modules, each containing an attention block and a feed-forward block, both with residual connections. Concurrently, ${S}^\mathrm{I}_{i}$ and ${S}^\mathrm{II}_i$ undergo feature-wise concatenation, where their relative positions in the feature dimension serve as implicit indicators distinguishing between the self and the partner's speech sources. The embeddings representing the denoising step $t$ and motion state $s$ are then summed with this concatenated representation. The resulting combined features are subsequently utilized to modulate the generation process through the AdaLN-Zero \cite{DiTiccv23} conditioning mechanism.

Our network does not explicitly incorporate motion history ${M}^\mathrm{I}_{i-1}$ as input. Instead, we leverage the tileable property of the diffusion model to maintain temporal coherence between consecutive motion segments during inference \cite{tseng2022edge}. Specifically, at the $i$-th round, we replace the initial portion of each $x_t$ with the terminal frames of ${M}^\mathrm{I}_{i-1}$, applying noise perturbation as defined in \Cref{equ:add_noise} to ensure consistency with the training setting.

\section{Data Process}
\label{appendix_data_process}
\subsection{Dataset Details}
With the increasing availability of conversational motion datasets, selecting appropriate and publicly accessible data is crucial for evaluating our system. Some recent datasets, such as GES-Inter \cite{qi2025co3gesturecoherentconcurrentcospeech} and the DND Group Gesture Dataset \cite{mughal2024convofusion}, are either not publicly available or do not conform to dyadic interaction scenarios. Therefore, we selected two high-quality, publicly available datasets: the Photoreal dataset \cite{ng2024audio2photoreal} and the InterAct dataset \cite{huang2024interact}.

The Photoreal dataset consists of approximately 8 hours of dyadic conversational data, including body and facial motion capture from four participants. It provides synchronized audio for both interlocutors but motion data for only one participant, encoded in a parametric format. To integrate this with our system, we used the authors’ official code to convert the parametric motion data into skeletal format (BVH). We used 2.5 hours of motion sequence data from speaker PXB for both training and evaluation, following the baseline setup in \cite{ng2024audio2photoreal}. The InterAct dataset includes roughly 8.3 hours of conversational interactions across daily-life scenarios, capturing separate motion and audio streams for each participant. It also includes frame-level annotations of motion states.

\subsection{Data Process}

To integrate the Photoreal \cite{ng2024audio2photoreal} and InterAct dataset \cite{huang2024interact} into our framework, we processed the motion and audio as follows:

\noindent\emph{Motion Processing.} For both datasets, we first applied mirror augmentation to the training data. We then segmented all motions into $5$-second clips, and translated each clip's starting point to the coordinate origin with orientation toward the forward direction (positive X-axis). Each clip consists of $150$ frames, corresponding to a frame rate of $30$ FPS. For pose representation at each frame, we used J = 57 joints for the Photoreal dataset and J = 48 joints for the InterAct dataset. Each joint was encoded using an exponential map representation. For the root joint, we employed absolute position and velocity relative to the previous frame. In summary, $M_t \in \mathbb{R}^{(J\times Q + G)} = \mathbb{R}^{(J\times 3 + 3 + 3)}$.

\noindent\emph{Audio Processing.} For audio processing, we applied the same segmentation approach as used for motion data. Inspired by \citet{DBLP:journals/tvcg/ZhangWLZLGJDGWL24}, we leveraged a pretrained WavLM \cite{wavlm} model to extract audio representations, as WavLM effectively captures complex and universal audio features.

\section{Details of User Study}
\label{appendix_user_study}

Our user experiments were conducted anonymously. For each test, participants watch two 10-second videos, each generated by different models (including the ground truth) for the same dyadic speech segment, played sequentially. The user study is conducted using the Human Behavior Online (HBO) tool provided by the Credamo platform \cite{credamo}. Participants are instructed to select their preferred video based on the provided evaluation criteria and rate their preference on a scale from 0 to 2, where 0 indicates no preference. The unselected video in the pair is assigned the inverse score (e.g., if a participant rates the chosen video 1, the other video receives -1).

For the \emph{human likeness} test, participants assess whether the generated motions resemble natural human movements. To eliminate potential bias from speech, these video clips are presented without audio. In the \emph{beat matching} test, participants evaluate the synchronization between the generated gestures and the speech rhythm. Since this metric primarily assesses single-person gestures, and a dyadic setting could introduce confounding factors, we render the videos with only one character's motion and corresponding audio for this evaluation. For the \emph{interaction level} test, participants determine whether the generated motions effectively convey dialogue interaction intentions between two individuals. To ensure participants clearly understand each evaluation criterion and can accurately distinguish between them, we provide detailed instructions as guidance:

\begin{itemize}
  \item \emph{Human Likeness:} Judge whether the generated gestures look natural and resemble real human movements. Focus on the smoothness, variety, and realism of body motions. Good examples should show natural transitions between gestures, avoiding excessive repetition or abrupt changes. Poor examples may appear stiff, mechanical, or contain unnatural jittering.
  \item \emph{Beat Matching:} Evaluate whether the gestures are synchronized with the rhythm of the speech. Check if gestures occur at appropriate times, matching emphases, pauses, or speech rhythm. Good examples align gestures with key words or stressed syllables. Poor examples may show gestures that lag, anticipate incorrectly, or are unrelated, resulting in poor coordination.
  \item \emph{Interaction Level:} Assess whether the two characters show signs of interaction. Good examples include mutual gaze (indicating attention and engagement), responsive actions such as nodding or imitating the partner’s gestures, appropriate movement toward or away from the partner, and natural physical contact when suitable. Poor examples show gaze avoidance, lack of mutual attention, or absence of responsive gestures, making the conversation feel disconnected.
\end{itemize}

For both the Photoreal and InterAct datasets, each participant completes 48 questions, each corresponding to a video pair, evenly divided into three categories: human likeness, beat matching, and interaction level tests. The experiment takes approximately 20 minutes to complete. We recruited 100 participants for each dataset via Credamo, resulting in a total of 200 participants. To ensure response validity, attention checks were embedded within each test category, and responses failing these checks were excluded from the final analysis. For statistical analysis, we conducted a one-way ANOVA followed by a post-hoc Tukey multiple comparison test for each user study. The assumptions of normality, homogeneity of variances, and independence were verified and met for all ANOVA tests.

\section{Details of DMSS Metric}
\label{appendix_dmss}
We propose Delayed Motion Synchrony Score (DMSS) to evaluate phase-shifted motion synchrony between two interacting agents. Given two joint velocity sequences, $M^\mathrm{I} \in \mathbb{R}^{T \times D}$ and $M^\mathrm{II} \in \mathbb{R}^{T \times D}$, DMSS computes the maximum Pearson correlation coefficient over a range of temporal frame shifts $\tau \in [-L, L]$, where L is the maximum allowable lag. The DMSS is formally defined as:

\begin{equation}
\text{DMSS}(M^\mathrm{I}, M^\mathrm{II}) = \max_{\tau \in [-L, L]} \rho (M^\mathrm{I}_{[\tau]}, M^\mathrm{II}_{[-\tau]} )
\end{equation}

where $\rho(\cdot, \cdot)$ denotes the Pearson correlation coefficient computed along the temporal dimension, and the shifted motion sequences $M_{[\tau]}$ and $M_{[-\tau]}$ are defined as:

\begin{equation}
M^{[\tau]} =
\begin{cases}
M[\tau:T] & \text{if } \tau > 0, \\
M[0:T+\tau] & \text{if } \tau < 0, \\
M[0:T] & \text{if } \tau = 0,
\end{cases}
\quad
M^{[-\tau]} =
\begin{cases}
M[0:T-\tau] & \text{if } \tau > 0, \\
M[-\tau:T] & \text{if } \tau < 0, \\
M[0:T] & \text{if } \tau = 0.
\end{cases}
\end{equation}

Prior to computing the correlation, both motion windows are z-score normalized to ensure scale invariance. Only upper-body joint velocities are used as input features, as they are more informative for capturing interactive motion cues. In our implementation, we use a window length $T = 30$ frames and a maximum lag $L = 5$. By definition, DMSS takes values in the range [-1, 1].

However, this metric has certain limitations. While a high DMSS indicates strong temporal synchrony, it does not distinguish between intentional coordination (e.g., mirroring or responsive gestures) and incidental motion similarity. Additionally, DMSS does not account for spatial interaction cues, such as the relative distance or orientation between the two agents, which are often crucial for capturing the nuances of interaction.

\section{Fine-grained Ablation of Dynamic Controller Agent}

To dissect the contribution of each component within our Dynamic Controller Agent (DCA), we conduct a fine-grained ablation study. Since DCA consists of three components—Gesture Synchrony Predictor, Spatial Relation Predictor, and Gaze Predictor—we ablate one component at a time while keeping the other two, and observe the effect on the generated behaviors. This results in three ablated variants: Ours (w/o Gesture Sync Predictor), Ours (w/o Spatial Relation Predictor), and Ours (w/o Gaze Predictor). We include these alongside two additional baselines: the full model and Ours (w/o DCA). Following the protocol in \Cref{subsubsec_user_study}, we generate gestures for ten audio segments on the Photoreal test set, and perform pairwise user comparisons to evaluate the Interaction Level metric. This design isolates the perceptual impact of each social signal.

\begin{table}[t]
\centering
\caption{Average scores of user study on the fine-grained ablation of DCA, with 95\% confidence intervals.}
\label{tab:ablation_granularity}
\resizebox{0.8\columnwidth}{!}{
\begin{tabular}{l c}
\hline
System & Interaction Level $\uparrow$ \\
\hline
Ours (w/o DCA) & -0.31 \\
Ours (w/o Gaze Predictor) & -0.05 \\
Ours (w/o Gesture Sync Predictor) & 0.02 \\
Ours (w/o Spatial Relation Predictor) & 0.13 \\
Ours & \textbf{0.22} \\
\hline
\end{tabular}
}
\end{table}

The results, presented in \Cref{tab:ablation_granularity}, clearly show that removing any single predictor degrades the perceived Interaction Level, confirming the positive contribution of all three DCA components. The Gaze Predictor's impact is the most pronounced; its removal causes the score to plummet from 0.22 to -0.05, resulting in a negative user preference. This underscores the critical role of gaze in conveying attention and engagement in social interactions. The Gesture Synchrony Predictor is the second most crucial component, followed by the Spatial Relation Predictor. This fine-grained analysis not only complements the baseline result of ablating the entire DCA module but also demonstrates that each signal plays a distinct and valuable role in generating high-quality dyadic social behaviors.

\section{Details of Prompt Ablation Experiment}
\label{appendix_prompt_ablation}

As shown in \Cref{appendix_Prompts}, our full prompt is carefully designed with four key components:
\begin{itemize}
  \item \emph{Reference Behavioral Theories:} Social and psychological principles drawn from linguistic and human behavior research, providing theoretical grounding for spatial and interactional reasoning.
  \item \emph{Mapping Rules:} Heuristic rules that translate qualitative spatial descriptions (e.g., “front-left”, “side-by-side”) into structured representations such as clock-based orientation and 2D movement vectors. These rules are applied specifically to spatial relation modeling.
  \item \emph{Task Definition:} A formally defined reasoning objective that instructs the agent to perform interaction analysis, along with explicit specifications of the input schema and output format.
  \item \emph{Stepwise Reasoning Guide:} An explicit chain-of-thought structure that guides the model through step-by-step spatial reasoning and decision-making.
\end{itemize}

We evaluate the quality of outputs under different prompt settings using the LLM-as-a-judge protocol~\cite{zhang2023wider}, which has been shown to strongly align with human judgments~\cite{zheng2023judgingllmasajudgemtbenchchatbot}.
Two evaluation criteria are defined: For the Scene Designer Agent, we visualize the predicted proxemic layout within its scene context and present it to the judging LLM, which assesses \emph{Layout Plausibility}—the plausibility and contextual fit of the spatial configuration. For the Dynamic Controller Agent, we provide the predicted interaction control signals along with the ongoing interaction context. The judging LLM evaluates \emph{Behavior Appropriateness}—whether the behavior aligns with the social and contextual expectations. We use GPT-4.1~\cite{openai2025gpt41} as the judging LLM, with explicit instructions to rate each output on a 1–5 scale. All evaluations are conducted independently to minimize bias and improve reliability.

\section{Diversity Analysis}

Diversity of generated gestures is an important aspect of naturalistic behavior modeling.
During qualitative evaluation, we observe that the generated gestures sometimes repeat the same actions, leading to limited behavioral variety.
To examine whether the constraints introduced by the Dynamic Controller Agent (DCA) reduce diversity, we compute the diversity metric ($\text{Div}_k$)~\cite{ng2024audio2photoreal} on the Photoreal test set. The results are shown in Table~\ref{tab:diversity}.

The results demonstrate that: (1) the diversity of our full model is close to that of ground truth; (2) including DCA does not reduce diversity—in fact, it slightly improves diversity compared to the version without DCA, likely because DCA encourages a broader range of interactive behaviors. This suggests that the primary limitation on diversity stems from the dataset itself, which contains only 2.5 hours of recordings from a single actor with a consistent speaking style.

\begin{table}[t]
\centering
\caption{Quantitative comparison of diversity scores ($\text{Div}_k$) on the Photoreal dataset. All systems are trained on the same dataset and evaluated using the same test audio inputs.}
\label{tab:diversity}
\resizebox{0.4\columnwidth}{!}{
\begin{tabular}{l c}
\hline
System & $\text{Div}_k$ \\
\hline
Ground Truth & 2.13 \\
\hline
Ours & \textbf{1.98} \\
Ours (w/o DCA) & 1.94 \\
LDA & 1.41 \\
\hline
\end{tabular}
}
\end{table}

\section{Extending Single-Person Gesture Generator}
To demonstrate the versatility of our Social Agent System, we can integrate it with a single-person gesture generation framework based on a diffusion-based architecture. The Social Agent System operates independently of the low-level gesture generator, enabling easy decoupling and integration into existing single-person models. As a case study, we incorporate our baseline model, GestureDiffuCLIP \cite{ao2023gesturediffuclip}, trained on the ZeroEGGS dataset \cite{ghorbani2022zeroeggs}. To enable dyadic interaction synthesis, we extend GestureDiffuCLIP by adding a dual-person auto-regressive inference strategy and incorporating interaction control signals through our Social Agent System. 

\begin{figure}[t]
  \centering
  \includegraphics[width=\columnwidth]{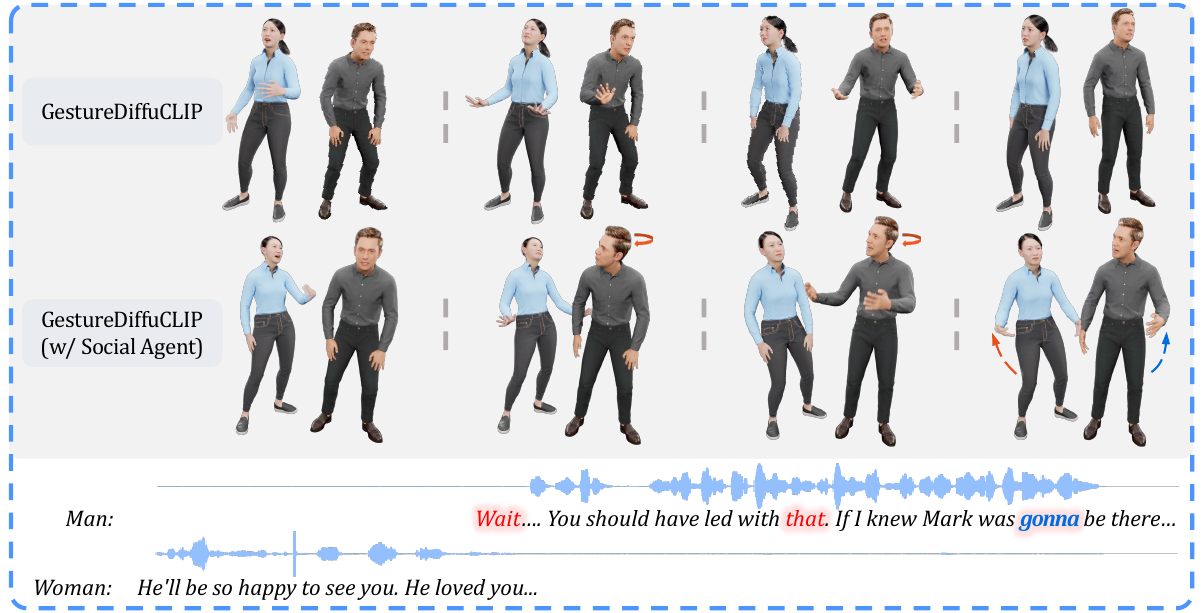}
  \caption{Comparison of GestureDiffuCLIP \cite{ao2023gesturediffuclip} outputs before and after incorporating our Social Agent System. The figure demonstrates how our framework enables a single-person gesture generator, originally lacking dyadic interaction capability, to synthesize realistic nonverbal behaviors between two characters, showcasing its effectiveness in interactive motion generation.}
  \label{fig:res_3}
\end{figure}

\Cref{fig:res_3} presents the visualization results comparing the original outputs of GestureDiffuCLIP \cite{ao2023gesturediffuclip} with those generated after integrating our Social Agent System. It can be observed that the original GestureDiffuCLIP model, as a single-person gesture generation model, lacks the capability to synthesize dyadic interactive behaviors. However, after integrating our Social Agent System, the model successfully generates interactive behaviors such as gaze and gesture imitation, significantly enhancing the realism of dyadic interactions. This integration effectively equips the model with the ability to generate coherent nonverbal interactions between two characters. These results further demonstrate the strength and scalability of our framework in enabling interactive behavior generation.

\section{Details of Spatial Relation Planner}
\label{appendix_SP}

In this section, we detail the implementation of the Spatial Relation Planner. We first introduce the classification of positional configurations. Next, we describe how the agent systematically converts qualitative spatial relationships into quantitative values using predefined mapping rules. Finally, we explain the process of global spatial calculation, where the predicted relative spatial information is transformed into global coordinates for motion initialization.

\subsection{Details of Positional Configuration}
\label{appendix_Positional}
\begin{figure}[t]
  \centering
  \includegraphics[width=\columnwidth]{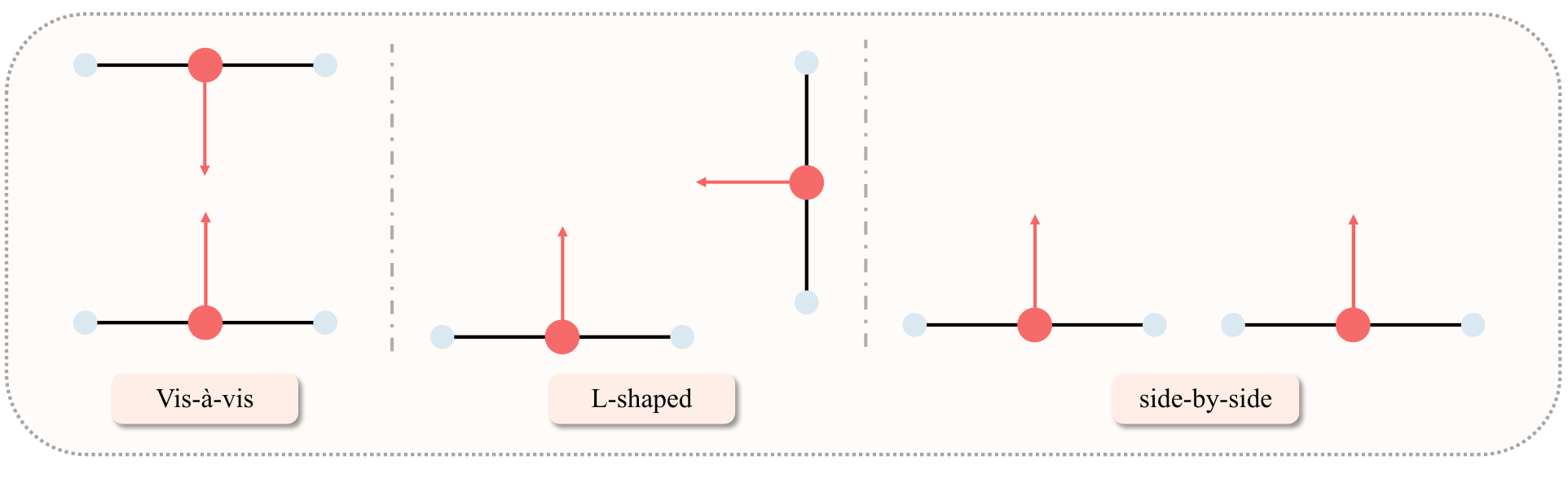}
  \caption{The three possible positional configurations in dyadic interactions, as described in Kendon's F-formation system \cite{kendon1990conducting}, are visualized in our diagram inspired by \cite{barua2021detectingsociallyinteractinggroups}.}
  \label{fig:fformation}
\end{figure}

As shown in \Cref{fig:fformation}, according to Kendon's F-formation system \cite{kendon1990conducting}, the positional configurations in dyadic conversations typically fall into one of the following three categories:
\begin{itemize}
    \item \emph{Vis-à-vis}: Both characters face each other directly;
    \item \emph{L-shaped}: Both characters are slightly angled toward one side;
    \item \emph{Side-by-Side}: Both characters stand shoulder to shoulder, facing the same direction.
    
\end{itemize}

\subsection{Mapping Rules for Quantitative Conversion}
\label{appendix_map}

As described in \Cref{sec:sda_32}, after obtaining qualitative results, the agent applies predefined mapping rules to convert positional relationships into quantitative values. The conversion process relies on three key relative parameters:
\begin{itemize}
    \item \emph{Direction $\theta$}: orientation of Character II with respect to Character I around the vertical axis;
    \item \emph{Direction  $\varphi$}: orientation of Character I with respect to Character II around the vertical axis;
    \item \emph{Distance $d$}: horizontal distance between them.
\end{itemize}

For relative directional values, the agent first translates the predicted positional configuration into relative directional descriptions (e.g., Character I is in front of Character II), which are then mapped to clock-based directional values (e.g., Character I is at Character II's 11:50 direction) for easier numerical computation.
For distance values, the agent selects an appropriate numerical distance based on the spatial distance category:
\begin{itemize}
    \item \emph{Interpersonal distance}: 0.5 – 0.7 meters;
    \item \emph{Social distance}: 0.7 – 1.2 meters;
    \item \emph{Public distance}: 1.2 – 2.0 meters.
\end{itemize}
These mapping rules are pre-defined and provided to the agent as guidelines, allowing it to predict the final numerical relative spatial values. These rules are also used in Spatial Relation Predictor. Detailed mapping rule prompts can be found in \Cref{appendix_srp_Prompts}.

\subsection{Global Spatial Information Calculation}
\label{appendix_proximic}

\begin{figure}[t]
  \centering
  \includegraphics[width=0.8\columnwidth]{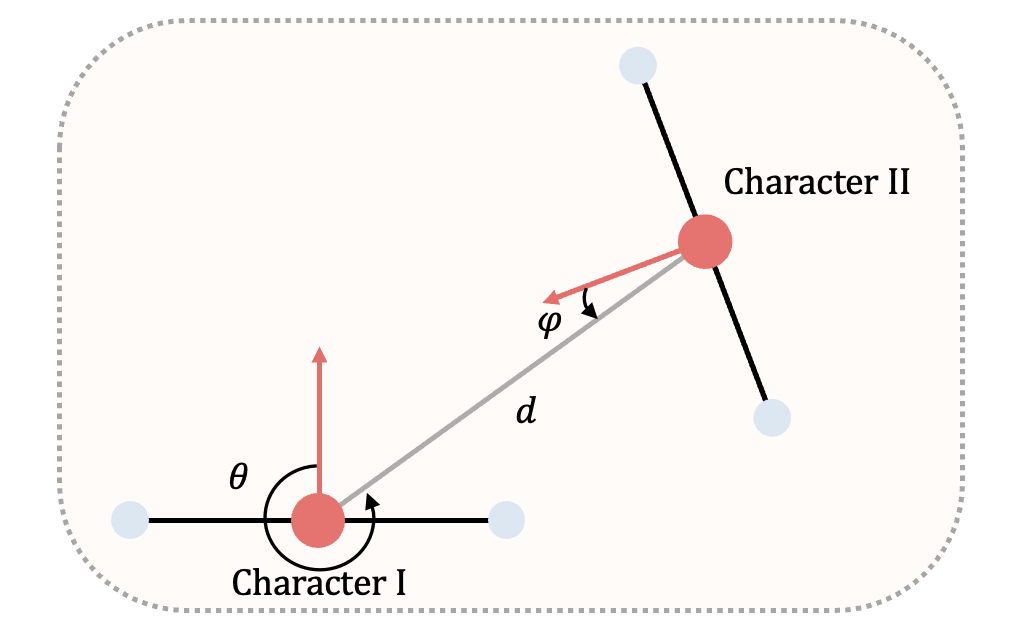}
  \caption{Representation of the relative spatial relationship between Character I and Character II from a top-down perspective.}
  \label{fig:relative_rep}
\end{figure}
This section details how we convert the relative spatial information predicted by the Spatial Relation Planner into global positions and orientations for motion initialization. 
As shown in \Cref{fig:relative_rep}, we first fix Character I's global horizontal position $\vec{p_{\mathrm{I}}}$ and orientation around the vertical axis $\alpha_{\mathrm{I}}$. Using the predicted relative parameters described in \Cref{appendix_map}: $\theta$, $\varphi$ and $d$, we could compute Character $\mathrm{II}$'s global horizontal position $\vec{p_{\mathrm{II}}}$ and orientation around the vertical axis $\alpha_{\mathrm{II}}$ as:

\begin{equation}
    \vec{p_\mathrm{II}} = \vec{p_\mathrm{I}} + d\begin{bmatrix} \cos(\alpha_\mathrm{I} + \theta) \\ \sin(\alpha_\mathrm{I} + \theta) \end{bmatrix}, \alpha_\mathrm{II} = \alpha_\mathrm{I} + \theta + \pi - \varphi
\end{equation}

This process determines the initial proxemic setup for motion generation. In our experiments, we set Character I's global horizontal position $\vec{p_{\mathrm{I}}}$ at [0, 0] and orientation around the vertical axis $\alpha_{\mathrm{I}}$ to 0.

\section{Social Agent prompts}
\label{appendix_Prompts}
As detailed in \Cref{sec:agent}, all modules within the Agent System are implemented using a prompt-based design approach, with carefully crafted prompts tailored for each module. To ensure a structured and consistent output, we employ the response\_format mechanism\footnote{https://platform.openai.com/docs/guides/structured-outputs}, enforcing adherence to a predefined JSON schema. Below, we provide examples of the designed prompts used in the Social Agent System.

\subsection{Gesture Sync Predictor}
\label{appendix_gesturesyn_Prompts}

\begin{framed}
\begin{lstlisting}[breaklines=true, breakindent=0pt]
You are an AI assistant with expertise in 3D spatial knowledge, psychology, and behavioral sciences, specializing in guiding gesture and motion generation for dyadic conversation scenarios. Your task is to analyze the need for synchronized interaction adjustments in a dyadic conversation over the next round (2.5 seconds) based on the provided input data and generate reasonable adjustment recommendations.

## @\textbf{Input Data}@:
1. Scene context: {}
2. Previous Round Motion Description: {}
3. Next Round Information:
   - Upcoming dialogue transcripts: {}

## @\textbf{Reference Theories}@:
In dyadic conversations, Gesture Synchrony, also known as Behavioral Synchrony, is a common phenomenon that helps individuals convey and interpret nonverbal signals. It consisting of two primary forms:

- Matching: Also known as the \emph{chameleon effect}, it refers to unconscious imitation of a partner's body gestures. This type of synchrony strengthens rapport and alignment between speakers.
- Meshing: Refers to real-time responsive feedback behaviors from the listener, such as nodding, facial expressions, or subtle head movements. In this context, we focus exclusively on nodding as the representative form of meshing, which plays a key role in regulating turn-taking and signaling active listening and engagement.

The occurrence of synchrony depends on several contextual and relational factors:

1. Matching (Gesture Imitation) is more likely in:
   - Cooperative, emotionally expressive, or informal settings.
   - Interactions involving romantic partners, family, or close friends.
   - Unequal power dynamics, where subordinates tend to imitate superiors.
   - Cases where the listener has visual access to the speaker's gestures.
2. Meshing (nodding) often occurs:
   - As a listener's response to the speaker's emotionally salient or affirming statements.
   - When showing understanding, empathy, or agreement.
   - With increased frequency in supportive or rapport-building contexts.
   - Unequal power dynamics, where subordinates tend to give feedback like nodding.
3. Synchrony is less likely in:
   - Conflictual, highly formal, or hierarchical settings with low emotional openness.
   - When spatial positioning obstructs visual perception of the partner's actions.

## @\textbf{Task}@:
Based on the provided input data, assess the likelihood and type of gesture synchrony that may occur in the next round. Follow the steps below:
1. Analyze the conversational context, interaction status, interpersonal relationship, and spatial positioning to determine whether the synchrony is likely to occur and which kind of synchrony will occur (gesture imitation or nodding).
2. Identify the roles of the two characters:
   - Determine which character is the initiator and which is the responder. For matching: who is the imitator and who is being imitated. For meshing: who is speaking and who is providing feedback.
3. Determine the most likely moment for gesture imitation to occur:
   - Identify the word or phrase in the upcoming transcript that is most likely to trigger the synchrony behavior.
   - Output this key word or phrase from the transcript.
\end{lstlisting}
\end{framed}

\subsection{Spatial Relation Predictor}
\label{appendix_srp_Prompts}
\begin{framed}
\begin{lstlisting}[breaklines=true, breakindent=0pt]
You are an AI assistant with expertise in 3D spatial knowledge, psychology, and behavioral sciences, specializing in guiding gesture and movement generation for dyadic conversation scenarios. Your task is to analyze the spatial positioning and orientation adjustments required for two interacting individuals in the next round (2.5 seconds) based on the provided input data and generate reasonable adjustment recommendations.

## @\textbf{Input Data}@:
1. Scene context: {}
2. Previous Round Motion Description: {}
3. Next Round Information:
   - Upcoming dialogue transcripts: {}

## @\textbf{Reference Theories}@:
1. Adjustments should align with real-world conversational behavior logic, considering the following factors:
   - Typical behaviors of speakers and listeners: e.g., listeners tend to turn their heads toward the speaker or make slight gestures to signal engagement.
   - Interpersonal relationships and contextual needs: e.g., closer physical proximity in intimate relationships versus greater distance between strangers.
   - Spatial plausibility: Adjustments should be realistic and logical according to human behavior.
   - Interactive motion cues: If clear interaction movements are observed, adjust both characters' positions and distances accordingly.
   
2. If there is a change in body orientation, it should generally be accompanied by a positional shift.
   - For example, if a character rotates left, they typically move slightly forward in that direction.
3. Positional Configuration:
   - Vis-à-vis: Both characters are directly facing each other.
   - L-shaped: Both characters are slightly angled towards one side.
   - Side-by-Side: Both characters stand shoulder to shoulder, facing the same direction.

## @\textbf{Mapping Rules}@:
1. Positional Configuration Mapping Rules
- Vis-à-vis:
   - Character B is directly in front of Character A and Character A is also directly in front of Character B.
- L-shaped:
   - If character B is in Character A's front-left, then Character A is in Character B's front-right or directly right.
   - If Character B is in Character A's front-right, then Character A is in Character B's front-left or directly left.
- Side-by-Side:
   - If Character B is to Character A's direct left, then Character A is to Character B's direct right.
   - If Character B is to Character A's direct right, then Character A is to Character B's direct left.
2. Direction Mapping Rules
Convert relative directional descriptions (e.g., front@-@right) into clock-based directional descriptions:
   - Front: 11:15 - 12:45
   - Front-right: 12:45 - 2:15
   - Right: 2:15 - 3:45
   - Back-right: 3:45 - 5:15
   - Back: 5:15 - 6:45
   - Back-left: 6:45 - 8:15
   - Left: 8:15 - 9:45
   - Front-left: 9:45 - 11:15
3. Movement Direction and Distance Mapping
   - Movement Directions:
      - Front-right: 0° - 45°
      - Back-right: 135° - 180°
      - Back-left: 180° - 225°
      - Front-left: 315° - 360°
   - Movement Distance:
      - Small step adjustment: 0.1 - 0.2 meters
      - Significant displacement: 0.2 - 0.4 meters
4. Numerical Conventions
   - If the relative positioning remains unchanged, then both orientation and position remain mostly stable.
   - Orientation changes should be minimal, typically within two adjacent clock directions.
   - Distance values should be converted to centimeters.
   - Clock values should range from 1 to 12, and minute values from 0 to 59.

## @\textbf{Task}@:
Based on the input data, analyze and output the following adjustments step by step for both individuals at the next round:
1. Overall Situation Analysis
   - Briefly analyzing the overall current situation. Consider any relevant contextual cues (e.g., tone, actions, stated intentions, implicit alignments) that may influence spatial relation .
2. Prediction next round Positional Configuration
   - Describe the current Positional Configuration of individuals and predict how it will evolve in the next round.
   - If no major contextual changes occur (e.g., no sudden shifts indicated in the dialogue), maintain the previous Positional Configuration.
3. Qualitative Analysis of Orientation Adjustments
   - Determine whether orientation adjustments are needed to achieve the predicted Positional Configuration.
   - If an adjustment is needed, identify where each character positions the other relative to themselves (e.g., front-left, back-right).
4. Quantitative Analysis of Orientation Adjustments
   - Convert the qualitative results into clock-based directional values [hour, minute].
   - Example format: If Character B is positioned at Character A's front-left, based on the direction mapping rules, Character B should be between 9:45 and 11:15. After further analysis, we determine that Character B is at 10:05.
5. Qualitative Analysis of Positional Adjustments
   - Determine whether positional movement is needed to achieve the predicted Positional Configuration.
   - If significant orientation changes occur, movement in the corresponding direction is likely necessary.
   - Analyze whether characters move closer or farther apart based on their current distance.
6. Quantitative Analysis of Positional Adjustments
   - Convert qualitative results into movement direction and distance using 2D vector representation:
   - Format: [horizontal angle (0-360°), movement distance (cm)]
   - If a character remains stationary, output [0.0, 0.0].
\end{lstlisting}
\end{framed}

\subsection{Gaze Predictor}
\label{appendix_gaze_Prompts}

\begin{framed}
\begin{lstlisting}[breaklines=true, breakindent=0pt]
You are an AI assistant with expertise in 3D spatial knowledge, psychology, and behavioral sciences, specializing in guiding head orientation and gaze direction adjustments for dyadic conversation scenarios. Your task is to analyze the head orientation and gaze focus adjustments needed for the next round (2.5 seconds) of a dyadic conversation based on the provided input data and generate reasonable adjustment recommendations.

## @\textbf{Input Data}@:
1. Scene context: {}
2. Previous Round Motion Description: {}
3. Next Round Information:
   - Upcoming dialogue transcripts: {}
   
## @\textbf{Reference Theories}@:
Various factors influence gaze focus behavior in dyadic conversations, including:
1. Interpersonal Closeness:
   - In intimate relationships, individuals tend to maintain prolonged eye contact as a sign of trust, affection, and sincerity.
   - In formal or unfamiliar relationships, eye contact is minimized to maintain distance or avoid excessive intimacy.
2. Conversation Context and Setting:
   - Interactive discussions requiring feedback involve more frequent gaze behavior to ensure mutual understanding.
   - In sensitive discussions (e.g., topics involving embarrassment, guilt, or conflict), gaze is often avoided as a self-protection mechanism.
3. Counterpart's Actions and Emotional Expressions:
   - Gaze direction is influenced by body language and emotional cues.
   - For instance, when the interlocutor makes an expressive hand gesture, gaze may naturally shift toward that specific body part (e.g., left or right hand).
4. Roles and Social Hierarchy:
   - Listeners tend to maintain gaze toward the speaker as a sign of engagement and respect.
   - Speakers' gaze behavior varies depending on interaction demands and listener feedback.
   - Social status and power dynamics also affect gaze duration: subordinates tend to look at superiors more frequently.
5. Individual Personality and Emotional State:
   - Introverted individuals tend to avoid prolonged gaze, while extroverts engage in more direct eye contact.
   - Emotional states such as anxiety or nervousness may lead to gaze avoidance, whereas relaxed and comfortable states encourage increased eye contact.
6. Positional Influence:
   - If Character A cannot see Character B through simple head rotation and requires full-body rotation, then head orientation adjustment is unnecessary.

## @\textbf{Task}@:
Based on the input data, analyze whether Characters A or B will adjust their head orientation and gaze direction within the next round. Specifically, complete the following steps:
1. Analyze the gaze focus direction and head orientation adjustments considering the factors listed above.
2. Determine whether each character needs to turn their head to:
   - Shift their gaze toward the interlocutor
   - Deliberately avoid eye contact
3. If gaze is required, estimate gaze duration based on the speaker/listener roles and conversation context. Classify the gaze duration into three categories:
   - Long gaze: 1.8 - 2.5 seconds (sustained eye contact)
   - Medium gaze: 1.0 - 1.8 seconds (moderate eye contact)
   - Short gaze: less than 1.0 seconds (brief glance)
4. If gaze is required, identify the specific word or phrase in the next time segment's transcript during which the character is most likely to shift gaze toward the partner.
   - Output a single word or phrase from the Upcoming dialogue transcripts that represents this moment.

\end{lstlisting}
\end{framed}